\documentclass[english,twocolumn,aps,prb]{revtex4}
\usepackage[]{fontenc}
\usepackage[latin1]{inputenc}
\usepackage{amsmath}
\usepackage{graphicx}

\makeatletter

\usepackage{epsfig}



\usepackage{babel}
\makeatother
\begin{document}

\title{Optical conductivity of a granular metal at not very low temperatures:
a path-integral approach}

\author{V. Tripathi}

\affiliation{Theory of Condensed Matter Group, Cavendish Laboratory, Department
of Physics, University of Cambridge, J. J. Thomson Avenue, Cambridge
CB3 0HE, United Kingdom}

\author{Y. L. Loh}

\affiliation{Department of Physics, Purdue University, 525 Northwestern Avenue,
West Lafayette, IN 47907-2036, U.S.A.}

\begin{abstract}
We study the finite-temperature optical conductivity $\sigma(\omega,T)$
of a granular metal using a simple model consisting of a array of
spherical metallic grains. It is necessary to include quantum tunneling
and Coulomb blockade effects to obtain the correct temperature dependence
of $\sigma$, and to consider polarization oscillations to obtain
the correct frequency dependence. We have therefore generalized the
Ambegaokar-Eckern-Schön (AES) model for granular metals to obtain
an effective field theory incorporating the polarization fluctuations
of the individual metallic grains. In the absence of intergrain tunneling,
the classical optical conductivity is determined by polarization oscillations
of the electrons in the grains, $\sigma(\omega)=-(ine^{2}f\omega/m)/(\omega^{2}-\omega_{r}^{2}-i|\omega|/\tau_{\text{grain}}),$
where $\omega_{r}=e\sqrt{(4\pi/3m)n}$ is the resonance frequency,
$\tau_{\text{grain}}^{-1}$ is the relaxation rate for electron motion
within the grain, and $f$ is the volume fraction occupied by the
grains. At finite intergrain tunneling, we find that $\sigma(\omega)=-(ine^{2}\omega f/m)/(\omega^{2}-\omega_{r}^{2}-i|\omega|/\tau_{\text{rel}})+\sigma_{\mathrm{AES}}(\omega,T),$
where $\tau_{\text{rel}}^{-1}$ is the total relaxation rate that
includes the intragrain relaxation rate $\tau_{\text{grain}}^{-1}$
as well as intergrain tunneling effects, and $\sigma_{\mathrm{AES}}(\omega,T)$
is the conductivity of the granular system from the AES model obtained
by ignoring polarization modes. We calculate the temperature and frequency
dependence of the intergrain relaxation time, $\Gamma(\omega,T)=\tau_{\text{rel}}^{-1}-\tau_{\text{grain}}^{-1},$
and find it is different from $\sigma_{\mathrm{AES}}(\omega,T).$
For small values of dimensionless intergrain tunneling conductance,
$g\ll1,$ the DC conductivity obeys an Arrhenius law, $\sigma_{\text{AES}}(0,T)\sim ge^{-E_{c}/T},$
whereas the polarization relaxation may even decrease algebraically,
$\Gamma(\omega,T)\sim(g/E_{c}^{2})[T^{2}+(\omega/2\pi)^{2}],$ when
$\omega,T\ll E_{c}.$ 
\end{abstract}
\maketitle

\section{Introduction}

An inhomogeneous mixture of metallic and insulating phases exhibits
a transition between bulk metallic and bulk insulating behavior. When
the volume fraction of metal is large, the composite material is a
{}``dirty metal'' containing isolated impurities; when the volume
fraction of metal is very small, it is a {}``dirty insulator''.
Between these two extremes, there is a third state consisting of large
($\sim100$\AA) metallic regions separated by insulating walls. Such
systems are called granular metals. Granularity can arise automatically;
for instance, electronic phase segregation has been directly observed
in the pseudogap phase of cuprate superconductors\cite{lang1} and
in two-dimensional electron gases in semiconductor heterostructures.\cite{zhitenev2000}
Granular metals can also be deliberately created by sputtering a metal
onto an insulating substrate,\cite{abeles1,simon87,gerber97} by lithographic
deposition of quantum dots, or by self-assembly of metal nanoparticles
coated with organic molecules. \cite{beverly02} Some of these methods
allow control of disorder.

Granular metals are very interesting as their transport properties
--- in particular, the DC conductivity --- cannot be explained by
simple extrapolation from the neighboring metallic or insulating phases.
Another probe of the metal-insulator transition is the optical (AC)
conductivity. In this paper we study the frequency and temperature
dependence of the optical conductivity of granular metals. We begin
by forming comparative and contextual links with existing literature
on transport in dirty metals, dirty insulators, and granular metals
themselves.

\subsection{Dirty metals}

A {}``dirty metal'' consists of impurities embedded in a metallic
host. The electronic states at the Fermi energy are delocalized throughout
the solid, giving a finite conductivity at zero temperature. Thermal
excitations are detrimental to charge transport, so the DC conductivity
has a {}``metallic'' temperature dependence ($d\sigma/dT<0$). At
very low temperatures, electron-electron interactions and quantum
coherence\cite{abrahams,Gorkov,finkelshtein83,castellani84,altshuler1}
can conspire to give {}``insulating'' corrections to conductivity
($d(\Delta\sigma)/dT>0$), which are usually weak.

The optical conductivity is well described by Drude theory, \begin{align}
\sigma_{\text{Drude}}(\omega) & =\frac{ne^{2}}{m}\frac{\tau_{\text{Drude}}}{1+i|\omega|\tau_{\text{Drude}}},\label{drudecond1}\end{align}
 where $n$ is the conduction electron density and $\tau_{\text{Drude}}$
is the relaxation time, which may be temperature-dependent. At high
frequencies the optical conductivity is dominated by electronic inertia,
$\sigma_{\text{Drude}}(\omega)\approx ne^{2}/im\omega.$ There are
small coherence corrections to the Drude result at low temperatures.

\subsection{Dirty insulators}

A {}``dirty insulator'' or {}``dirty semiconductor'' consists
of impurities embedded in an insulating host. There is a finite density
of states at the Fermi energy due to impurity states, but these states
are all localized, so the DC conductivity is zero at $T=0$. Conduction
occurs by thermally activated hopping between bound states, so the
conductivity has an {}``insulating'' temperature dependence ($d\sigma/dT>0$);
it obeys a variable-range-hopping law of the Mott\cite{mottBook}
or Efros-Shklovskii\cite{efros75} kind depending on whether the long-range
Coulomb interaction is screened. In this paper we will not be studying
the effects of long-range Coulomb interaction, and therefore, we discuss
below only the Mott case. For the sake of completeness, a discussion
of the Efros-Shklovskii case is provided in Appendix \ref{sec:app0}.

Mott\cite{mott1970} showed that the main contribution to optical
conductivity comes from resonant absorption by pairs of states, one
of which is occupied and the other empty. Mott's argument, which we
recapitulate briefly, is valid when electron correlations due to long-range
Coulomb interactions can be disregarded. Let the two states in a pair
have energies $\epsilon_{i}$ and $\epsilon_{j}.$ The resonance condition
is satisfied when $\omega=\epsilon_{j}-\epsilon_{i}.$ The transition
rate $P_{ij}$ in presence of an electric field $E\cos(\omega t)$
is given by the Fermi Golden Rule, \begin{align}
P_{ij} & =\pi e^{2}{\mathcal{V}}|x_{ij}|^{2}E^{2}\rho(\epsilon_{j}),\label{goldenrule}\end{align}
 where $\rho(\epsilon)$ is the density of (impurity band) states
per unit volume and ${\mathcal{V}}$ is the volume of the system.
The conductivity $\sigma(\omega)$ is then found by multiplying by
$\omega/\frac{{\mathcal{V}}}{2}E^{2},$ averaging over all occupied
initial states with energies in the interval $\epsilon_{F}-\omega$
and $\epsilon_{F},$ and averaging over all unoccupied final states
$j.$ The result is\cite{mott1970}\begin{align}
\text{Re}(\sigma(\omega)) & =2\pi e^{2}{\mathcal{V}}\omega\int_{\epsilon_{F}-\omega}^{\epsilon_{F}}d\epsilon\,|x_{ij}|_{\text{av}}^{2}\rho(\epsilon)\rho(\epsilon+\omega).\label{mott1}\end{align}
 The best scenario for a hopping transition between the two localized
states at low frequencies is that they are degenerate and the splitting
of the levels to to tunneling, $\Delta\epsilon_{ij}\sim we^{-x_{ij}/\xi_{loc}}$
is smaller than $\omega,$ or in other words, the distance $x_{ij}$
between the localized states should be large enough: $x_{ij}\geq r_{\omega}=\xi_{loc}\ln(w/\omega).$
Here $w$ is an energy scale of the order of the relaxation rate.\cite{shklovskii81}
Localized states in a {}``shell'' of thickness $\xi_{loc}$ around
$r_{\omega}$ will also satisfy the condition for resonance. Using
this in Eq.(\ref{mott1}), we arrive at Mott's optical conductivity
for a disordered insulator, \begin{align}
\text{Re}(\sigma(\omega)) & \sim2\pi e^{2}n_{\text{imp}}^{2}(\omega/\delta)^{2}(r_{\omega}^{d-1}\xi_{loc})r_{\omega}^{2}\nonumber \\
 & \approx2\pi e^{2}n_{\text{imp}}^{2}(\omega/\delta)^{2}\xi_{loc}^{d+2}\ln^{d+1}(w/\omega),\label{mott2}\end{align}
 where $n_{\text{imp}}$ is the number density of localized states,
$d$ is the dimensionality, and $1/\delta$ is the density of states
at an impurity site. An important assumption in obtaining Eq.(\ref{mott2})
is that there is no inelastic scattering during the hopping process.

At high frequencies, $\omega\gg w,$ the electrons are not localized,
and the optical conductivity reverts to the Drude expression, Eq.(\ref{drudecond1}),
with $n_{\text{imp}}$ as the conduction electron density. At some
intermediate frequency, the optical conductivity has a maximum; however,
this maximum is just due to a crossover between different behaviors,
and is not associated with any special resonance.

\subsection{Granular metals}

A granular metal consists of metallic grains embedded in an insulating
host. The electrons are localized within each grain due to the Coulomb
blockade. Conduction occurs by intergrain tunneling of thermally excited
charges, so the DC conductivity has an insulating temperature dependence
($d\sigma/dT>0$). However, a granular metal differs from a dirty
insulator in that there is a large number of states $N=(a_{B}/R)^{-d}$
on each grain, so the mean level spacing $\delta\sim\epsilon_{F}/N$
is very small. For temperatures (or frequencies) higher than $\delta,$
these closely-spaced levels may be treated as a continuum leading
to incoherent or dissipative transport phenomena.\cite{ambegaokar82,girvin90,devoret90,panyukov91,beloborodov01,efetov02,arovas,altland04,lohgranular05}
Inelastic cotunneling, in particular, is the core of the variable-range-cotunneling
mechanism of charge transport in a disordered granular metal,\cite{feigelman05,beloborodov05}
and has an even greater effect on heat transport\cite{tripathi05}.
The low-energy particle-hole excitations within each grain also give
rise to a metallic linear-in-$T$ specific heat.

The standard model for studying dissipative transport in granular
superconductors was obtained by Ambegaokar, Eckern, and Schön (AES)
in 1982.\cite{ambegaokar82} This model has also been widely used
to study normal granular metals.\cite{girvin90,devoret90,panyukov91,beloborodov01,efetov02,arovas,altland04,lohgranular05}
It describes the competition between incoherent intergrain tunneling
(characterized by the dimensionless intergrain conductance $g$) that
tends to delocalize charge, and Coulomb blockade (characterized by
the charging energy of the grain, $E_{c}$) that suppresses intergrain
tunneling. These are quantum effects that are beyond the realm of
classical electrodynamics and circuit theory. The AES approach is
valid at temperatures larger than both the mean level spacing in a
grain $\delta$ and the Thouless energy of intergrain diffusion. \cite{beloborodov01,efetov02}
In this regime, intergrain transport is incoherent and quantum interference
effects are unimportant.

We now turn to optical conductivity. The study of optical properties
of metal particles has a long history and occupies a large body of
literature. \cite{Gorkov2,strassler1972,cohen1973,wood1982,kreibigBook,henning99}
Effective-medium theories are perhaps the most common approaches.\cite{stroud1975}
The earliest of these is due to Maxwell Garnett who in 1904 proposed
using frequency-dependent dielectric functions in the expression for
the effective dielectric constant of the granular metal that had been
obtained from electrostatics.\cite{garnett1904} Thus if $\varepsilon_{m}(\omega)$
and $\varepsilon_{i}(\omega)$ are the bulk dielectric functions of
the metallic and insulating phases, and $\varepsilon_{\mathrm{eff}}(\omega)$
is the effective dielectric constant of the composite,\begin{align}
\frac{\varepsilon_{\mathrm{eff}}(\omega)-\varepsilon_{i}(\omega)}{\varepsilon_{\mathrm{eff}}(\omega)+2\varepsilon_{i}(\omega)} & =f\frac{\varepsilon_{m}(\omega)-\varepsilon_{i}(\omega)}{\varepsilon_{m}(\omega)+2\varepsilon_{i}(\omega)},\label{maxwellgarnett}\end{align}
 where $f$ is the volume fraction of the metal. Alternatively, following
Bruggeman,\cite{bruggeman1935} one can treat the granular system
as a fraction $f$ of metal and $1-f$ of insulator immersed in an
effective medium. The effective dielectric function is obtained by
solving \begin{align}
f\frac{\varepsilon_{m}(\omega)-\varepsilon_{\mathrm{eff}}(\omega)}{\varepsilon_{m}(\omega)+2\varepsilon_{\mathrm{eff}}(\omega)}+(1-f)\frac{\varepsilon_{i}(\omega)-\varepsilon_{\mathrm{eff}}(\omega)}{\varepsilon_{i}(\omega)+2\varepsilon_{\mathrm{eff}}(\omega)} & =0.\label{bruggeman}\end{align}

In 1908 Mie recognized the importance of polarization oscillations
for the optical conductivity.\cite{mie1908} The classical optical
conductivity of a clean spherical metallic grain can be inferred from
the equation of motion of the electrons. Suppose an external field
$\mathbf{E}_{ext}e^{i\omega t}$ acts on a spherical metallic particle
and induces a polarization $\mathbf{P}.$ From classical electrodynamics,
the field $\mathbf{E}_{int}$ inside the particle is $\mathbf{E}_{int}=\mathbf{E}_{ext}-(4\pi/3)\mathbf{P}.$
Using the equation of motion of the electrons, $-\omega^{2}\mathbf{x}_{\omega}=-e\mathbf{E}_{int}/m,$
together with the definition of the current density $\mathbf{j}_{\omega}=-ine\omega\mathbf{x}_{\omega}$
and its relation to the polarization, $\mathbf{j}_{\omega}=i\omega\mathbf{P},$
and the external electric field, $\mathbf{j}_{\omega}(\mathbf{q}=0)=\sigma(\omega)\mathbf{E}_{ext},$
we arrive at \begin{align}
\sigma(\omega) & =-\frac{ine^{2}f}{m}\frac{\omega}{\omega^{2}-\omega_{r}^{2}},\label{sphereOptical}\\
\omega_{r}^{2} & =\frac{4\pi}{3}\frac{ne^{2}}{m}.\label{resonanceFreq}\end{align}
 $\omega_{r},$ the frequency of resonant polarization oscillations,
is smaller than the plasma frequency of the bulk metal, $e\sqrt{4\pi n/m}$,
by a factor of $1/\sqrt{3},$ and depends on the shape of the grain
but not on its size. %
\footnote{This paper deals with the `conduction resonance' at which $\mathrm{Re~}\sigma(\omega)$
is maximum, not the `plasma resonance' at which $\mathrm{Im~}\epsilon(\omega)^{-1}$
is maximum. See Ref.~\onlinecite{marton1971} for an insightful
discussion of both phenomena.%
} At very high frequencies, $\omega\gg\omega_{r},$ the optical conductivity
approaches that of a free particle, because the inertia of the electrons
prevents them from screening the external electric field. If the electrons
in the grain have a finite relaxation time, the equation of motion
$(-\omega^{2}+i|\omega|/\tau_{\text{rel}})\mathbf{x}_{\omega}=-e\mathbf{E}_{int}/m$
gives \begin{align}
\sigma(\omega) & =-\frac{ine^{2}}{m}\frac{\omega}{\omega^{2}-\omega_{r}^{2}-i|\omega|/\tau_{\text{rel}}}.\label{sphereOptical2}\end{align}

\begin{figure}
\includegraphics[scale=0.8]{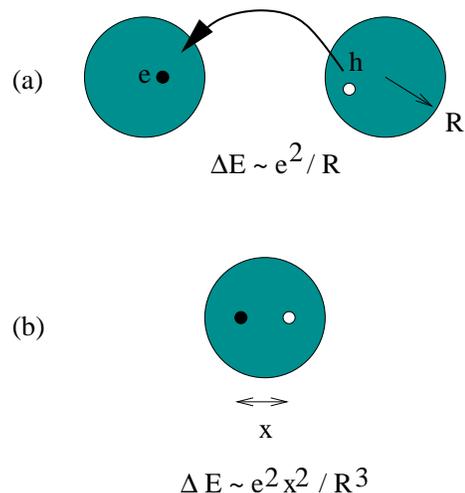}

\caption{\label{f:monopoledipole} (a) Polarization in the standard AES model
involves charge asymmetry between \emph{different} grains. For weak
intergrain tunneling, the energy $\Delta E$ associated with the polarization
is of the order of the charging energy of the grain, $e^{2}/R.$ (b)
Polarization due to uneven distribution of charge within a grain.
Such polarization excitations cost less energy than case (a) but are
not considered in the standard AES approach. The filled large circles
denote grains, and $e$ and $h$ denote electron excess and deficit,
respectively.}
\end{figure}

The Mie approach is entirely classical. In order to capture the temperature
dependence of $\sigma$, which is determined by tunneling and charging
effects, one has to use a quantum treatment such as AES effective-field
theory. In the original AES model, the Coulomb interaction is approximated
by a capacitance matrix; the electrostatic potential is uniform on
each grain (although it may fluctuate in time). This amounts to assuming
that the electrons are massless and can instantaneously redistribute
to suppress potential variations within the grain. Such a {}``monopole''
approximation is adequate insofar as DC transport is concerned, because
the bottleneck in transport is intergrain tunneling rather than electronic
inertia. Optical properties, however, depend crucially upon the finite
mass of the electrons and the possible polarization of individual
grains (see Fig.\ref{f:monopoledipole}). Indeed, a calculation of
$\sigma(\omega)$ from the AES action alone misses the polarization
resonance peak completely, and thus severely violates the sum rule.

\subsection{Purpose of this paper and results}

\begin{figure}
\includegraphics[scale=0.8]{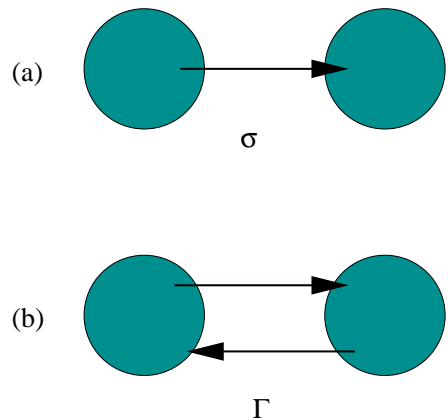}

\caption{\label{f:difference1} Physically different mechanisms involving
intergrain tunneling processes that determine the DC conductivity
$\sigma$ (that measures the escape rate of an electron from the grain)
and the relaxation rate of polarization oscillations $\Gamma.$ (a)
Coulomb blockade of intergrain tunneling, especially at small values
of $g,$ suppresses electron escape from the grains leading to $\sigma\sim\exp(-E_{c}/T)$
at low temperatures. (b) A polarization oscillation does not result
in a net transfer of charge and therefore no strong Coulomb blockade
at low temperatures. High temperatures and strong intergrain tunneling
both wash out Coulomb blockade effects, whereby the two processes
show the same temperature dependence. Drude theory predicts the same
temperature dependence for $\sigma$ and $\Gamma$.}
\end{figure}

In this paper, we generalize the Ambegaokar-Eckern-Schön (AES) model
for a regular array of spherical grains to include dipole (polarization)
as well as monopole (charge) degrees of freedom. Using this new effective
field theory, we are able to calculate the conductivity as a function
of temperature as well as of frequency:

\begin{enumerate}
\item Using a Kubo formula, we find that the optical conductivity of isolated
grains is mainly due to intragrain dipole oscillations,\begin{align*}
\sigma(\omega,T) & =-\frac{ine^{2}f}{m}\frac{\omega}{\omega^{2}-\omega_{r}^{2}-i|\omega|/\tau_{\text{grain}}},\end{align*}
 where $\tau_{\text{grain}}^{-1}$ is the relaxation rate for intragrain
scattering and consists of, apart from the classical Drude relaxation
in a bulk metal, additional finite-volume effects such as Landau damping.\cite{kawabata1966,wood1982,kreibig1985} 
\item At finite intergrain tunneling, we find that there is a small additional
{}``monopole'' contribution $\sigma_{\mathrm{AES}}(\omega,T)$ due
to intergrain charge oscillations, and that intergrain tunneling also
imparts an extra width $\Gamma$ to the dipole resonance:\begin{align*}
\sigma(\omega) & \approx\sigma_{\mathrm{AES}}(\omega,T)-\frac{ine^{2}f}{m}\frac{\omega}{\omega^{2}-\omega_{r}^{2}-i|\omega|/\tau_{\text{rel}}},\end{align*}
 where $\tau_{\text{rel}}^{-1}=\tau_{\text{grain}}^{-1}+\Gamma(\omega,T).$
At finite temperature, $\sigma_{\mathrm{AES}}(0,T)$ is finite and
gives the DC conductivity of the granular array. $\Gamma(\omega,T)$
depends on the intergrain dimensionless conductance, $g,$ and the
grain charging $E_{c}.$ It is independent of $\tau_{\text{grain}}^{-1}$,
and has a different temperature dependence. 
\item The temperature and frequency dependence of the resonance width $\Gamma(\omega,T)$
is different from $\sigma_{\mathrm{AES}}(\omega,T),$ especially when
$\omega,T$ are smaller than the effective charging energy of the
grains. At large $\omega,T,$ both $\Gamma$ and $\sigma_{\mathrm{AES}}$
become independent of $\omega,T$ and are proportional to the dimensionless
intergrain tunneling conductance, $g.$ Fig.\ref{f:difference1} illustrates
the physical difference between the two. The qualitative difference
in the manner in which intergrain tunneling affects $\sigma_{\mathrm{AES}}$
and $\Gamma$ cannot be explained by a simple effective medium approximation.\cite{landauer1952} 
\item The optical conductivity of a granular metal is physically different
from a dirty insulator with Coulomb interaction even though both show
similar features. For both systems, $\sigma(\omega)$ vanishes as
a power law at low and high frequencies and has a maximum at intermediate
frequencies. However, for a granular metal this maximum is due to
resonant polarization oscillations, whereas for a dirty insulator
the maximum is just due to a crossover and is not associated with
any resonance. 
\end{enumerate}

Our theory neglects the interactions between dipole degrees of freedom
on different grains. However, at high frequencies (so that the metal
dielectric function approaches unity) or for small $f,$ our approximation
approaches the Maxwell Garnett result, Eq.(\ref{maxwellgarnett}),
when we use the standard relation $\mathop{\mathrm{Re}}\sigma(\omega)=-\frac{|\omega|}{4\pi}\mathop{\mathrm{Im}}\varepsilon_{\mathrm{eff}}(\omega).$
In many experimental situations, such as two-dimensional granular
arrays, it is possible to screen out long-range Coulomb interaction
within the sample by using a gate electrode, in which case an effective
medium treatment is not necessary. In any case, the dipole-dipole
interactions can be modeled with a matrix if necessary, just as the
monopole-monopole interactions are included in the original AES model
as the capacitance matrix. For simplicity, we also ignore possible
effects arising from the non-uniformity of the shape of the metallic
particles.

\subsection{Arrangement of the paper}

In Sec.\ref{sec:Model} we introduce our model of granular metals:
an array of spherical metallic grains with interacting electrons and
a finite intergrain hopping. We then make a multipole expansion of
the potential on the grain in terms of spherical harmonics up to the
$l=1$ (dipole) component. In Sec.\ref{sec:Effective-field} we develop
an effective field theory in terms of the monopole and dipole components
of the potential fluctuations: this is a generalization of the AES
theory of transport in granular metals. Optical conductivity is calculated
in Sec.\ref{sec:Optical-conductivity}. First we consider isolated
grains and calculate the optical conductivity using the Kubo formula
approach as well as from the dielectric function. The calculation
with the dielectric function is much less tedious. The result agrees
with classical expressions for the optical conductivity of isolated
grains. Next we consider the case of finite intergrain tunneling and
study the differences from the classical optical conductivity. An
explicit expression for the broadening of the polarization resonance,
$\Gamma,$ is obtained. Its temperature and frequency dependence are
found to be different from the conductivity of the granular metal
obtained from the AES model. The paper concludes (Sec.\ref{sec:Conclusions})
with a discussion of the results and open problems for further study.

\section{Model \label{sec:Model}}

We consider the following action for the granular metal array, \begin{align}
S & =\frac{e^{2}}{2}\sum_{\mathbf{ij}}\int_{\tau\mathbf{x_{i}}\mathbf{x}_{j}}\rho(\mathbf{x_{i}},\tau)\rho(\mathbf{x_{j}},\tau)\frac{1}{|\mathbf{x_{i}}-\mathbf{x_{j}}|}+\nonumber \\
 & +\sum_{\mathbf{i}}\int_{\tau\mathbf{x_{i}}}\psi_{\tau\mathbf{x_{i}}}^{\dagger}[\partial_{\tau}+\xi(-i\nabla_{\mathbf{x}_{\mathbf{i}}})]\psi_{\tau\mathbf{x_{i}}}+\nonumber \\
 & +\sum_{\langle\mathbf{ij}\rangle}\int_{\tau\mathbf{x_{i}}\mathbf{x}_{\mathbf{j}}}t_{\mathbf{x_{i}},\mathbf{x}_{\mathbf{j}}}\psi_{\tau\mathbf{x_{i}}}^{\dagger}\psi_{\tau\mathbf{x}_{\mathbf{j}}},\label{hamilt1}\end{align}
 where $\rho(\mathbf{x_{i}},\tau)=(\psi_{\tau\mathbf{x_{i}}}^{\dagger}\psi_{\tau\mathbf{x_{i}}\tau}-Q_{0\mathbf{i}})\Theta(|\mathbf{x_{i}}|-R)$
is the excess electronic charge density at position $\mathbf{x_{i}}$
in the $\mathbf{i}^{th}$ grain of radius $R,$ $\xi(-i\nabla_{\mathbf{i}})=\epsilon(-i\nabla_{\mathbf{x}_{\mathbf{i}}})-\mu=\mathbf{p}_{\mathbf{x}_{\mathbf{i}}}^{2}/2m-\mu,$
$\mathbf{a}$ is a lattice translation vector, and $t_{\mathbf{x_{i}},\mathbf{x_{j}}}$
is the intergrain hopping amplitude. Integrals over $\tau$ are understood
to go from 0 to $\beta$. We assume the intergrain hopping amplitude
has a white-noise distribution,\begin{align}
\langle t_{\mathbf{x_{i}},\mathbf{x}_{\mathbf{j}}}t_{\mathbf{x_{k}},\mathbf{x}_{\mathbf{l}}}\rangle & =|t|^{2}\delta(\mathbf{x_{i}}-\mathbf{x_{l}})\delta(\mathbf{x_{j}}-\mathbf{x_{k}}).\label{whitenoise}\end{align}
 where angle brackets denote disorder averaging.

Next we decouple the Coulomb interaction in Eq.(\ref{hamilt1}) through
a Hubbard-Stratonovich field, $V_{\mathbf{i}}(\mathbf{x_{i}})$, that
has the physical meaning of the electrostatic potential. The interaction
part of the action is \begin{align}
S_{int} & =-\frac{1}{2e^{2}}\sum_{\mathbf{ij}}\int_{\tau\mathbf{x_{i}}\mathbf{x_{j}}}\!\!\!\!\! C(\mathbf{x_{i}},\mathbf{x_{j}})V_{\mathbf{i}}(\mathbf{x_{i}},\tau)V_{\mathbf{j}}(\mathbf{x_{j}},\tau)\nonumber \\
 & +\sum_{\mathbf{i}}\int_{\tau\mathbf{x_{i}}}\psi_{\tau\mathbf{x_{i}}}^{\dagger}V_{\mathbf{i}}(\mathbf{x_{i}},\tau)\psi_{\tau\mathbf{x_{i}}},\label{hubbard}\end{align}
 where $\int_{\mathbf{x_{j}}}C(\mathbf{x_{i}},\mathbf{x_{j}})\frac{1}{|\mathbf{x_{j}}-\mathbf{x_{k}}|}=\delta(\mathbf{x_{i}}-\mathbf{x_{k}})$
subject to appropriate boundary conditions at the metallic grains;
thus $C(\mathbf{x_{i}},\mathbf{x_{j}})$ is proportional to the Laplace
operator.

For simplicity, we will consider grains sufficiently far apart so
that the mutual interaction of electrons on different grains is small
compared to the interaction of electrons within individual grains.
With this simplification, the interaction part of the action becomes\begin{align}
S_{int} & \approx-\frac{1}{8\pi e^{2}}\sum_{\mathbf{i}}\int_{\tau\mathbf{x}}\,\mathbf{E_{i}}(\mathbf{x},\tau)\cdot\mathbf{E_{i}}(\mathbf{x},\tau)+\nonumber \\
 & +\sum_{\mathbf{i}}\int_{\tau\mathbf{x_{i}}}\psi_{\tau\mathbf{x_{i}}}^{\dagger}V_{\mathbf{i}}(\mathbf{x_{i}},\tau)\psi_{\tau\mathbf{x_{i}}},\label{hubbard2}\end{align}
 $\mathbf{E_{i}}(\mathbf{x},\tau)=-\nabla V_{\mathbf{i}}(\mathbf{x},\tau)$
is the electric field at $\mathbf{x}$ due to charge on an isolated
grain at $\mathbf{i}.$ The potential away from the boundary may be
expanded in a basis of eigenfunctions of the Laplace equation, \begin{align*}
V_{\mathbf{i}}(\mathbf{x},\tau) & =\sum_{lm}A_{\mathbf{i}}^{lm}(\tau)\left(\frac{r}{R}\right)^{l}Y_{lm}(\theta,\phi)\,\,\,\,(r<R),\\
V_{\mathbf{i}}(\mathbf{x},\tau) & =\sum_{lm}B_{\mathbf{i}}^{lm}(\tau)\left(\frac{R}{r}\right)^{l+1}Y_{lm}(\theta,\phi)\,\,\,\,(r>R),\end{align*}
 where $r=|\mathbf{x}|$. Continuity of $V_{\mathbf{i}}(\mathbf{x},\tau)$
at the boundary requires that $A_{\mathbf{i}}^{lm}(\tau)=B_{\mathbf{i}}^{lm}(\tau).$
For the purposes of this paper, it is sufficient to retain just the
monopole component (average potential) \begin{align*}
V_{\mathbf{i}}(l=0;\mathbf{x},\tau) & =\left\{ \begin{array}{r}
V_{\mathbf{i}0}(\tau),\qquad r<R\\
V_{\mathbf{i}0}(\tau)(R/r),\qquad r>R,\end{array}\right.\end{align*}
 and the dipole components (electric field),\begin{align*}
V_{\mathbf{i}}^{(\alpha)}(l=1;\mathbf{x},\tau) & =\left\{ \begin{array}{r}
V_{\mathbf{i}1}^{(\alpha)}(\tau)(x^{\alpha}/R),\qquad r<R\\
V_{\mathbf{i}1}^{(\alpha)}(\tau)(R^{2}x^{\alpha}/r^{3}),\qquad r>R.\end{array}\right.\end{align*}
 Using the definition $\mathbf{E_{i}}(\mathbf{x},\tau)=-\nabla V_{\mathbf{i}}(\mathbf{x},\tau)$
the interaction part of the action, Eq.(\ref{hubbard2}) takes the
form \begin{align}
S_{int}\approx & -\frac{R}{2e^{2}}\sum_{\mathbf{i}}\int_{\tau}\left[(V_{\mathbf{i}0}(\tau))^{2}+\sum_{\alpha}(V_{\mathbf{i}1}^{(\alpha)}(\tau))^{2}\right]\nonumber \\
+ & \sum_{\mathbf{i}}\int_{\tau\mathbf{x_{i}}}\!\!\!\psi_{\tau\mathbf{x_{i}}}^{\dagger}\left[V_{\mathbf{i}0}+\sum_{\alpha}V_{\mathbf{i}1}^{(\alpha)}(x_{\mathbf{i}}^{\alpha}/R)\right]\psi_{\tau\mathbf{x_{i}}}.\label{multipole}\end{align}
 Now make the gauge transformations\begin{align}
\psi_{\tau\mathbf{x_{i}}}\rightarrow\psi_{\tau\mathbf{x_{i}}}\, e^{-i\varphi_{\mathbf{i}0}(\tau)-i\sum_{\alpha}\varphi_{\mathbf{i}1}^{(\alpha)}(\tau)(x_{\mathbf{i}}^{\alpha}/R)},\label{gauge}\\
V_{\mathbf{i}0}(\tau)=i\partial_{\tau}\varphi_{\mathbf{i}0},\,\, V_{\mathbf{i}1}^{(\alpha)}(\tau)=i\partial_{\tau}\varphi_{\mathbf{i}1}^{(\alpha)}.\label{gauge2}\end{align}
 to eliminate $V_{\mathbf{i}0}(\tau)$ and replace $V_{\mathbf{i}1}(\tau)$
by a time-dependent vector potential,\begin{align}
S_{el} & =\sum_{\mathbf{i}}\int_{\tau\mathbf{x_{i}}}\psi_{\tau\mathbf{x}_{\mathbf{i}}}^{\dagger}\bigg[\partial_{\tau}+\xi(-i\nabla_{\mathbf{x}_{\mathbf{i}}})+V_{\mathbf{i}0}+\nonumber \\
 & +\sum_{\alpha}V_{\mathbf{i}1}^{(\alpha)}(x_{\mathbf{i}}^{\alpha}/R)\bigg]\psi_{\tau\mathbf{x_{i}}}\nonumber \\
 & \rightarrow\sum_{\mathbf{i}}\!\!\int_{\tau\mathbf{x_{i}}}\!\!\!\!\psi_{\tau\mathbf{x_{i}}}^{\dagger}\left[\partial_{\tau}+\xi(-i\nabla_{\mathbf{x}_{\mathbf{i}}}-\pmb\varphi_{\mathbf{i}1}/R)\right]\psi_{\tau\mathbf{x_{i}}}.\label{gauge3}\end{align}
 The gauge transformations also dress the tunneling element in Eq.(\ref{hamilt1})
with monopole ($l=0$) and dipole ($l=1$) phase fluctuations,\begin{align}
t_{\mathbf{x_{i}},\mathbf{x}_{\mathbf{j}}}\rightarrow\tilde{t}_{\mathbf{x_{i}},\mathbf{x}_{\mathbf{j}}}(\tau)=t_{\mathbf{x_{i}},\mathbf{x}_{\mathbf{j}}}e^{i(\varphi_{\mathbf{i}0}(\tau)-\varphi_{\mathbf{j}0}(\tau))}\times\nonumber \\
\times\exp\left[(i/R)\sum_{\alpha}\left(\varphi_{\mathbf{i}1}^{(\alpha)}(\tau)x_{\mathbf{i}}^{\alpha}-\varphi_{\mathbf{j}1}^{(\alpha)}(\tau)x_{\mathbf{j}}^{\alpha}\right)\right].\label{gauge4}\end{align}

\section{Effective field theory \label{sec:Effective-field}}

Integrating out the conduction electrons results in an effective action
for the $l=0$ and $l=1$ phase fluctuations:\begin{align}
S_{\mathrm{eff}}[\varphi] & =\frac{R}{2e^{2}}\sum_{\mathbf{i}}\int_{\tau}\left[\left(\partial_{\tau}\varphi_{\mathbf{i}0}\right)^{2}+\left|\partial_{\tau}\pmb\varphi_{\mathbf{i}1}\right|^{2}\right]\nonumber \\
- & \text{tr ln}\left[G_{\pmb\varphi_{\mathbf{i}1}}^{-1}\delta_{\mathbf{ij}}-\left(\tilde{t}_{\mathbf{x_{i}},\mathbf{x_{j}}}\delta_{\mathbf{j},\mathbf{i}+\mathbf{a}}+\mathbf{i}\leftrightarrow\mathbf{i}+\mathbf{a}\right)\right],\label{seff1}\end{align}
 where \begin{align}
-G_{\pmb\varphi_{\mathbf{i}1}}^{-1} & =\partial_{\tau}+\frac{1}{2m}\left(\mathbf{p}_{\mathbf{x}_{\mathbf{i}}}-\frac{\pmb\varphi_{\mathbf{i}1}}{R}\right)^{2}-\mu\label{Ginv}\end{align}
 is the inverse of the electron Green function on grain $\mathbf{i}$
in the absence of intergrain tunneling, and $\tilde{t}_{\mathbf{x}_{\mathbf{i}},\mathbf{x}_{\mathbf{j}}}$
is the dressed tunneling amplitude defined in Eq.(\ref{gauge4}) and
the bare tunneling $t_{\mathbf{x}_{\mathbf{i}},\mathbf{x}_{\mathbf{j}}}$
has a gaussian distribution as in Eq.(\ref{whitenoise}). We study
first the effective field theory for isolated grains and then consider
the effect of finite intergrain tunneling.

\subsection{Isolated grains}

In the absence of tunneling, the {}``bare'' effective action $S_{\mathrm{eff}}^{(0)}[\varphi]$
is obtained by expanding the determinant in Eq.(\ref{seff1}) up to
second order in $\pmb\varphi_{\mathbf{i}1}(\tau)$, \begin{align}
S_{\mathrm{eff}}^{(0)}[\varphi] & =\frac{R}{2e^{2}}\sum_{\mathbf{i}}\int_{\tau}\left[\left(\partial_{\tau}\varphi_{\mathbf{i}0}\right)^{2}+\left|\partial_{\tau}\pmb\varphi_{\mathbf{i}1}\right|^{2}\right]\nonumber \\
 & +\frac{1}{2mR^{2}}\sum_{\mathbf{i}}\left[\int_{\tau\mathbf{x_{I}}}G_{\mathbf{i}}^{(0)}(\mathbf{x_{i}},\mathbf{x_{i}};\tau,\tau)\pmb\varphi_{\mathbf{i}1}^{2}(\tau)\right.\nonumber \\
 & +\frac{1}{m}\int_{\tau\tau'\mathbf{x_{i}x_{i}}'}\pmb\varphi_{\mathbf{i}1}(\tau)\cdot\mathbf{p}_{\mathbf{x}_{\mathbf{i}}}\, G_{\mathbf{i}}^{(0)}(\mathbf{x_{i}},\mathbf{x}'_{\mathbf{i}};\tau,\tau')\nonumber \\
 & \times\pmb\varphi_{\mathbf{i}1}(\tau')\cdot\mathbf{p}_{\mathbf{x}_{\mathbf{i}}'}\, G_{\mathbf{i}}^{(0)}(\mathbf{x}'_{\mathbf{i}},\mathbf{x_{i}};\tau',\tau)\bigg],\label{seff2}\end{align}
 where $G_{\mathbf{i}}^{(0)}=-\left[\partial_{\tau}+\frac{1}{2m}\mathbf{p}_{\mathbf{x}_{\mathbf{i}}}^{2}-\mu\right]^{-1}$
is the bare electron Green function, \begin{align}
G_{\mathbf{i}}^{(0)}(\mathbf{x}_{\mathbf{i}},\mathbf{x}_{\mathbf{i}}';\tau,\tau') & =T\sum_{\lambda,n}\frac{\psi_{\lambda}(\mathbf{x}_{\mathbf{i}}')\psi_{\lambda}^{*}(\mathbf{x}_{\mathbf{i}})}{i\nu_{n}-\xi_{\mathbf{i}\lambda}}e^{-i\nu_{n}(\tau-\tau')}\nonumber \\
 & =\sum_{\lambda}G_{\mathbf{i}\lambda}^{(0)}(\tau,\tau')\psi_{\lambda}(\mathbf{x}_{\mathbf{i}}')\psi_{\lambda}^{*}(\mathbf{x}_{\mathbf{i}}).\label{baregreen}\end{align}
 Note that\begin{align}
\sum_{\lambda}G_{\mathbf{i}\lambda}^{(0)}(\tau,\tau') & \approx\nu(\epsilon_{F})T\sum_{n}\int_{-\tau_{c}^{-1}}^{\tau_{c}^{-1}}d\xi_{\mathbf{i}}\frac{e^{-i\nu_{n}(\tau-\tau')}}{i\nu_{n}-\xi_{\mathbf{i}}}\nonumber \\
 & =-2i\nu(\epsilon_{F})T\sum_{n}e^{-i\nu_{n}(\tau-\tau')}\cot^{-1}\nu_{n}\tau_{c}\label{localgreenfn2}\end{align}
 where $\tau_{c}\sim\epsilon_{F}^{-1}$ is a short time cutoff. For
$|\tau-\tau'|\gg\tau_{c},$ this simplifies to\begin{align}
\sum_{\lambda}G_{\mathbf{i}\lambda}^{(0)}(\tau,\tau') & \approx\frac{\pi T\nu(\epsilon_{F})}{\sin\pi T(\tau-\tau')}.\label{localgreenfn}\end{align}
 We shall use this expression unless stated otherwise.

Eq.(\ref{seff2}) can be presented in a more recognizable form as
\begin{align}
S_{\mathrm{eff}}^{(0)}[\varphi] & =\frac{R}{2e^{2}}\sum_{\mathbf{i}}\int_{\tau}\left[\left(\partial_{\tau}\varphi_{\mathbf{i}0}\right)^{2}+\left|\partial_{\tau}\pmb\varphi_{\mathbf{i}1}\right|^{2}\right]\nonumber \\
+ & \frac{4\pi}{3}\frac{R}{2e^{2}}\sum_{\mathbf{i}\alpha\beta}\int_{\tau,\tau'}K_{\mathbf{i}}^{\alpha\beta}(\tau-\tau')\varphi_{\mathbf{i}1}^{(\alpha)}(\tau)\varphi_{\mathbf{i}1}^{(\beta)}(\tau').\label{seff3}\end{align}
 in terms of the bare electromagnetic response function of the $\mathbf{i}^{th}$
grain, \begin{align}
K_{\mathbf{i}}^{\alpha\beta}(\tau-\tau')=\delta^{\alpha\beta}\frac{e^{2}}{Vm}\int_{\mathbf{x_{i}}}G_{\mathbf{i}}^{(0)}(\mathbf{x_{i}},\mathbf{x_{i}};\tau,\tau')\delta(\tau-\tau')+\nonumber \\
\!\!\!\!\!\!+\frac{e^{2}}{Vm^{2}}\int_{\mathbf{x_{i}x}'_{\mathbf{i}}}\!\!\!\!\!\! p_{\mathbf{x}_{\mathbf{i}}}^{\alpha}G_{\mathbf{i}}^{(0)}(\mathbf{x_{i}},\mathbf{x}'_{\mathbf{i}};\tau,\tau')p_{\mathbf{x}_{\mathbf{i}}'}^{\beta}G_{\mathbf{i}}^{(0)}(\mathbf{x}'_{\mathbf{i}},\mathbf{x_{i}};\tau',\tau),\,\,\,\,\label{emr1}\end{align}
 where $V=(4\pi/3)R^{3}$ is the volume of the grain. In a bulk metal,
the two terms in the electromagnetic response function in Eq.(\ref{emr1})
would correspond to the diamagnetic and paramagnetic parts of the
bulk conductivity $\sigma(\omega)=K(\omega)/i\omega$. In a finite
system, the situation is trickier. The frequency dependence of $K$
is \begin{align}
K_{\mathbf{i}}^{\alpha\beta}(i\omega_{m})=\delta^{\alpha\beta}\frac{ne^{2}}{m}\left[1+\frac{2}{mN}\sum_{\lambda\lambda'}\frac{f(\xi_{\mathbf{i}\lambda})\xi_{\mathbf{i},\lambda\lambda'}|p_{\mathbf{i},\lambda\lambda'}^{\alpha}|^{2}}{\omega_{m}^{2}+\xi_{\mathbf{i},\lambda\lambda'}^{2}}\right],\label{emr2}\end{align}
 where $\lambda,\lambda'$ label the eigenvalues of the free electron
Hamiltonian of a grain, $\xi_{\mathbf{i},\lambda\lambda'}=\xi_{\mathbf{i}\lambda}-\xi_{\mathbf{i}\lambda'},$
$f(\xi)=[e^{\beta\xi}+1]^{-1}$ is the Fermi-Dirac distribution function,
and $p_{\mathbf{i},\lambda\lambda'}^{\alpha}=\langle\lambda|p_{\mathbf{i}}^{\alpha}|\lambda'\rangle.$
$n=N/V$ is the number density of electrons in a grain. If the temperature
(or frequency) is much smaller than the level separation $\delta=\xi_{\lambda\lambda'}^{avg}\sim\epsilon_{F}/N,$
we expand the right hand side of Eq.(\ref{emr2}) in ascending powers
of $\omega_{m}:$\begin{align}
K_{\mathbf{i}}^{\alpha\beta}(i\omega_{m})\approx\delta^{\alpha\beta}\frac{ne^{2}}{m}\bigg[1+\frac{2}{mN}\sum_{\lambda\lambda'}\frac{f(\xi_{\mathbf{i}\lambda})|p_{\mathbf{i},\lambda\lambda'}^{\alpha}|^{2}}{\xi_{\mathbf{i},\lambda\lambda'}}\nonumber \\
-\frac{2}{mN}\omega_{m}^{2}\sum_{\lambda\lambda'}\frac{f(\xi_{\mathbf{i}\lambda})|p_{\mathbf{i},\lambda\lambda'}^{\alpha}|^{2}}{\xi_{\mathbf{i},\lambda\lambda'}^{3}}\bigg]+O(\omega_{m}^{4}).\label{emrsmallf}\end{align}
 The static part of Eq.(\ref{emrsmallf}) can be shown to vanish using
the Reiche-Thomas-Kuhn sum rule,\cite{reiche1925,kuhn1925}\begin{align}
\frac{2}{m}\sum_{\lambda'}\frac{|p_{\mathbf{i},\lambda\lambda'}^{\alpha}|^{2}}{\xi_{\mathbf{i},\lambda\lambda'}} & =-1,\label{trksumrule}\end{align}
 along with the identity $\sum_{\lambda}f(\xi_{\mathbf{i}\lambda})=N.$
Combining Eq.(\ref{seff3}) and Eq.(\ref{emrsmallf}), one finds that
the surviving contribution in Eq.(\ref{emrsmallf}) makes a finite
size quantum correction to the RPA dielectric constant\cite{Gorkov2,strassler1972,cohen1973,wood1982,kreibigBook,henning99},
$\varepsilon^{RPA}=1-\frac{2e^{2}}{m^{2}R^{3}}\sum_{\lambda\lambda'}\frac{f(\xi_{\mathbf{i}\lambda})|p_{\mathbf{i},\lambda\lambda'}^{\alpha}|^{2}}{\xi_{\mathbf{i},\lambda\lambda'}^{3}}\sim1-(k_{F}a_{0})(R/a_{0})^{2},$
where $a_{0}$ is a small length of the order of a lattice constant.
As a result, even for metallic grains a few tens of lattice constants
across, the static dielectric constant rapidly approaches bulk values
(where is it infinity), and the polarizability, $\alpha=R^{3}(\varepsilon^{RPA}-1)/(\varepsilon^{RPA}+2),$
approaches the classical value, $\alpha_{\text{classical}}=R^{3}.$

The sum rule enables us to recast the electromagnetic response function
as \begin{align}
K_{\mathbf{i}}^{\alpha\beta}(i\omega_{m}) & =-\delta^{\alpha\beta}\frac{2e^{2}}{Vm^{2}}\omega_{m}^{2}\!\sum_{\lambda\lambda'}\frac{f(\xi_{\mathbf{i}\lambda})|p_{\mathbf{i},\lambda\lambda'}^{\alpha}|^{2}}{\xi_{\mathbf{i},\lambda\lambda'}(\omega_{m}^{2}+\xi_{\mathbf{i},\lambda\lambda'}^{2})},\label{emr3}\end{align}
 which is a known result. For the rest of the paper, unless stated
otherwise, we shall assume that the temperature (or frequency) is
much larger than the level separation $(T/\delta\gg1)$. Then, using
Eq.(\ref{trksumrule}) and Eq.(\ref{emr3}), we obtain \begin{align}
K_{\mathbf{i}}^{\alpha\beta}(i\omega_{m}) & \approx\delta^{\alpha\beta}\frac{ne^{2}}{m},\qquad T/\delta\gg1,\label{emr4}\end{align}
 that is, the value for the clean bulk metal; this is the diamagnetic
response due to electron acceleration in an electric field. Hence,
Eq.(\ref{seff3}) becomes \begin{align}
S_{\mathrm{eff}}^{(0)}[\varphi] & =\frac{R}{2e^{2}}\sum_{\mathbf{i}}\int_{\tau}\left[\left(\partial_{\tau}\varphi_{\mathbf{i}0}\right)^{2}+\left|\partial_{\tau}\pmb\varphi_{\mathbf{i}1}\right|^{2}+\omega_{r}^{2}\left|\pmb\varphi_{\mathbf{i}1}\right|^{2}\right]\label{seffbare}\end{align}
 where $\omega_{r}$ is the resonance frequency for a metallic sphere,
\begin{align}
\omega_{r}^{2} & =\frac{4\pi}{3}\frac{ne^{2}}{m},\end{align}
 we introduced in Eq.(\ref{resonanceFreq}). In a collisionless bulk
metal, the paramagnetic part of the electromagnetic response function
defined in Eq.(\ref{emr1}) vanishes. However, in a finite-size grain,
if one can treat the quasiparticle excitations in the grain as a continuum
(this is so if the temperature is not too low, $T\gg\delta$), the
paramagnetic part is finite and gives rise to a finite relaxation
of the oscillations through disintegration into incoherent particle-hole
excitations.\cite{kawabata1966,wood1982} The relaxation time has
been shown in numerous works\cite{kawabata1966,wood1982,kreibig1985}
to be of the order of the time of flight, $R/v_{F}.$ Physically,
the relaxation is due to Landau damping of plasma oscillations at
a finite wavevector: the minimum wavevector in a grain of size $R$
is of the order of $\pi/R.$ Other inelastic processes such as phonon
scattering will also contribute to relaxation.

\subsection{Finite intergrain tunneling}

We now obtain the effective field theory when intergrain tunneling
is finite. At not too low temperatures,\cite{beloborodov01,efetov02}
$T\gg\text{max }(|t|^{2}\delta,\delta),$ and for large enough\cite{zarand2000}
grains $(k_{F}R)^{2}\gg1,$ it suffices to expand the electron determinant
in Eq.(\ref{seff1}) up to $O(t^{2})$, \begin{align}
 & S_{\mathrm{eff}}^{\mathrm{tun}}[\varphi]=\frac{1}{2}\sum_{\mathbf{i},\mathbf{a}}\int_{\tau\tau'\mathbf{x_{i}}\mathbf{x}_{\mathbf{i}}'\mathbf{x}_{\mathbf{i}+\mathbf{a}}\mathbf{x}_{\mathbf{i}+\mathbf{a}}'}\!\!\!\!\tilde{t}_{\mathbf{x}_{\mathbf{i}}',\mathbf{x}_{\mathbf{i}+\mathbf{a}}'}(\tau')\tilde{t}_{\mathbf{x_{\mathbf{i}+\mathbf{a}}},\mathbf{x_{i}}}(\tau)\nonumber \\
 & \times G_{\pmb\varphi_{\mathbf{i}1}}(\mathbf{x}_{\mathbf{i}},\mathbf{x}'_{\mathbf{i}};\tau,\tau')G_{\pmb\varphi_{\mathbf{i}+\mathbf{a},1}}(\mathbf{x}'_{\mathbf{i}+\mathbf{a}},\mathbf{x}_{\mathbf{\mathbf{i}+\mathbf{a}}};\tau',\tau)\nonumber \\
 & =\frac{|t|^{2}}{2}\sum_{\mathbf{i},\mathbf{a}}\int_{\tau\tau'\mathbf{x_{i}}\mathbf{x}_{\mathbf{i}+\mathbf{a}}}G(\mathbf{x}_{\mathbf{i}},\mathbf{x}{}_{\mathbf{i}};\tau,\tau')G_{\pmb\varphi_{\mathbf{i}+\mathbf{a},1}}(\mathbf{x}{}_{\mathbf{i}+\mathbf{a}},\mathbf{x}_{\mathbf{\mathbf{i}+\mathbf{a}}};\tau',\tau)\nonumber \\
 & e^{i(\varphi_{\mathbf{ij},0}(\tau')-\varphi_{\mathbf{ij},0}(\tau))}\nonumber \\
 & e^{i(1/R)(\pmb\varphi_{\mathbf{i}1}(\tau')-\pmb\varphi_{\mathbf{i}1}(\tau))\cdot\mathbf{x_{i}}}e^{-i(1/R)(\pmb\varphi_{\mathbf{j}1}(\tau')-\pmb\varphi_{\mathbf{j}1}(\tau))\cdot\mathbf{x}_{\mathbf{j}}},\label{stunn1}\end{align}
 where $\varphi_{\mathbf{ij},0}(\tau)=\varphi_{\mathbf{i}0}(\tau)-\varphi_{\mathbf{j}0}(\tau).$
The $\pmb\varphi_{\mathbf{i}1}$ dependence in Eq.(\ref{stunn1})
comes from the exponential as well as from the the Green functions,
$G_{\pmb\varphi_{\mathbf{i}1}}(\mathbf{x}_{\mathbf{i}},\mathbf{x}_{\mathbf{i}};\tau,\tau'),$
etc. We show in Appendix \ref{sec:app1} that the contribution arising
from the expansion of $G_{\pmb\varphi_{\mathbf{i}1}}(\mathbf{x}_{\mathbf{i}},\mathbf{x}_{\mathbf{i}};\tau,\tau')$
etc. in powers of $\pmb\varphi_{\mathbf{i}1}$ is insignificant compared
to that coming from the exponential. Therefore we expand only the
exponential in Eq.(\ref{stunn1}) up to second order in $\pmb\varphi_{\mathbf{i}1}$
etc. and ignore the $\pmb\varphi_{\mathbf{i}1}$ dependence of the
Green functions. Thus the tunneling part of the effective action is
\begin{align}
S_{\mathrm{eff}}^{\mathrm{tun}}[\varphi] & \approx\!\!|t|^{2}\!\!\sum_{\mathbf{i},\mathbf{a};\lambda\lambda'}\!\int_{\tau\tau'}\!\!\!\!\Pi_{\mathbf{i},\mathbf{i}+\mathbf{a}}(\tau,\tau')G_{\mathbf{i}\lambda}^{(0)}(\tau,\tau')G_{\mathbf{i}+\mathbf{a},\lambda'}^{(0)}(\tau',\tau)\nonumber \\
- & \frac{|t|^{2}}{6R^{2}}\!\!\sum_{\mathbf{i},\mathbf{a};\lambda\lambda'}\int_{\tau\tau'}\!\!\!\!\!\!\Pi_{\mathbf{i},\mathbf{i}+\mathbf{a}}(\tau,\tau')G_{\mathbf{i}\lambda}^{(0)}(\tau,\tau')G_{\mathbf{i}+\mathbf{a},\lambda'}^{(0)}(\tau',\tau)\nonumber \\
\times & \big[r_{\mathbf{i}\lambda}^{2}(\pmb\varphi_{\mathbf{i}1}(\tau')-\pmb\varphi_{\mathbf{i}1}(\tau))^{2}\nonumber \\
 & +r_{\mathbf{i}+\mathbf{a},\lambda'}^{2}(\pmb\varphi_{\mathbf{i}+\mathbf{a},1}(\tau')-\pmb\varphi_{\mathbf{i}+\mathbf{a},1}(\tau))^{2}\big]\label{stunn2}\end{align}
 where\begin{align}
\Pi_{\mathbf{i},\mathbf{i}+\mathbf{a}}(\tau,\tau') & =\cos(\varphi_{\mathbf{i},\mathbf{i}+\mathbf{a},0}(\tau)\!-\!\varphi_{\mathbf{i},\mathbf{i}+\mathbf{a},0}(\tau'))\label{ctaudef}\end{align}
 and $r_{\mathbf{i}\lambda}^{2}=\langle\mathbf{i}\lambda|\hat{r}^{2}|\mathbf{i}\lambda\rangle$
are the matrix elements of $\hat{r}^{2}=|\mathbf{\hat{\mathbf{x}}}|^{2}$
with the eigenstates of grain $\mathbf{i}$. For a spherical grain,
the eigenfunctions are spherical Bessel functions $j_{n}(\kappa_{nl}r/R)Y_{lm}(\theta,\phi),$
where $\kappa_{nl}$ is the $l^{th}$ zero of $j_{n}(x).$ Numerically
evaluating the matrix elements we find they range between $0.28R^{2}$
and $R^{2}.$ In particular, $\lim_{n\rightarrow\infty}(r^{2})_{n1}=R^{2},$
and $\lim_{l\rightarrow\infty}(r^{2})_{nl}=R^{2}/3.$ Thus the matrix
elements of $r^{2}$ do not vary strongly and are of the order of
$R^{2}.$ Hence Eq.(\ref{stunn2}) may be written \begin{align}
S_{\mathrm{eff}}^{\mathrm{tun}}[\varphi]\approx-\frac{\pi gT^{2}}{2}\sum_{\mathbf{i},\mathbf{a}}\int_{\tau\tau'}\frac{\Pi_{\mathbf{i},\mathbf{i}+\mathbf{a}}(\tau,\tau')}{\sin^{2}\pi T(\tau-\tau')}\nonumber \\
+\frac{\pi gT^{2}b}{2}\sum_{\mathbf{i},\mathbf{a}}\int_{\tau\tau'}\frac{\Pi_{\mathbf{i},\mathbf{i}+\mathbf{a}}(\tau,\tau')}{\sin^{2}\pi T(\tau-\tau')}\nonumber \\
\times\big[(\pmb\varphi_{\mathbf{i}1}(\tau')-\pmb\varphi_{\mathbf{i}1}(\tau))^{2}+(\pmb\varphi_{\mathbf{i}+\mathbf{a},1}(\tau')-\pmb\varphi_{\mathbf{i}+\mathbf{a},1}(\tau))^{2}\big];\qquad\label{stunn3}\end{align}
 here $b\sim0.1$ is a constant, $g$ is the dimensionless intergrain
tunneling conductance \begin{align}
g & =2\pi|t|^{2}\nu(\epsilon_{F})^{2},\label{g}\end{align}
 Eqs.(\ref{seffbare}) and (\ref{stunn3}) form the effective action,
$S_{\mathrm{eff}}[\varphi]=S_{\mathrm{eff}}^{(0)}[\varphi]+S_{\mathrm{eff}}^{\mathrm{tun}}[\varphi]$,
which generalizes the AES action to include the physics of dipolar
oscillations. This may be presented as $S_{\mathrm{eff}}[\varphi]\approx S_{\mathrm{AES}}[\varphi_{0}]+S_{pol}[\varphi_{0}]$
where\begin{align}
 & S_{\mathrm{pol}}[\varphi]\approx\nonumber \\
 & \frac{T}{2(e^{2}/R)}\sum_{\mathbf{i},m}(\omega_{m}^{2}+\omega_{r}^{2})\pmb\varphi_{\mathbf{i}1}(\omega_{m})\cdot\pmb\varphi_{\mathbf{i}1}(-\omega_{m})+\nonumber \\
 & \frac{T}{4(e^{2}/R)}\sum_{\mathbf{i}\mathbf{a}}\Gamma_{\mathbf{i},\mathbf{i}+\mathbf{a}}(i\omega_{m})|\omega_{m}|\times\nonumber \\
 & [\pmb\varphi_{\mathbf{i}1}(\omega_{m})\cdot\pmb\varphi_{\mathbf{i}1}(-\omega_{m})+\pmb\varphi_{\mathbf{i}+\mathbf{a},1}(\omega_{m})\cdot\pmb\varphi_{\mathbf{i}+\mathbf{a},1}(-\omega_{m})],\label{Spol}\end{align}
 where \begin{align}
S_{\mathrm{AES}}[\varphi_{0}] & =\frac{1}{2(e^{2}/R)}\sum_{\mathbf{i}}\int_{\tau}(\partial_{\tau}\varphi_{\mathbf{i}0})^{2}+\nonumber \\
 & -\frac{\pi gT^{2}}{2}\sum_{\mathbf{ia}}\int_{\tau\tau'}\frac{\Pi_{\mathbf{i},\mathbf{i}+\mathbf{a}}(\tau,\tau')}{\sin^{2}\pi T(\tau-\tau')}\qquad\label{SAES}\end{align}
 is the standard Ambegaokar-Eckern-Schön (AES) model for normal granular
metals and\begin{align}
\Gamma_{\mathbf{i},\mathbf{i}+\mathbf{a}}(i\omega_{m})=\frac{e^{2}4\pi gbT^{2}}{R|\omega_{m}|}\!\!\int_{\tau}(1-e^{i\omega_{m}\tau})\frac{\Pi_{\mathbf{i},\mathbf{i}+\mathbf{a}}(\tau,0)}{\sin^{2}(\pi T\tau)}.\label{Gamma1}\end{align}

The quantities $\Pi$ and $\Gamma$ are functionals of $\varphi_{0}$.
In principle, fluctuations of $\varphi_{0}$ and of $\varphi_{1}$
can influence each other since they both appear in $S\mathrm{_{\mathrm{pol}}}$.
In practice, it is sufficient to calculate the correlator $\langle\Pi[\varphi_{0}]\rangle$
for $S\mathrm{_{\mathrm{\mathrm{AES}}}}$ alone, and to use this mean
value in $S\mathrm{_{\mathrm{pol}}}$ to determine the fluctuations
of $\varphi_{1}$. To justify this, we show that fluctuations of $\varphi_{1}$
have a negligible effect on the {}``kernel'' for $\varphi_{0}$
. If in Eq.(\ref{stunn3}) we average over the fields $\pmb\varphi_{\mathbf{i}1}$
using their bare propagator in the absence of tunneling (see Eq.(\ref{seffbare})),
\begin{align}
\langle(\pmb\varphi_{\mathbf{i}1}(\tau)-\pmb\varphi_{\mathbf{i}1}(0))^{2}\rangle & =3(e^{2}/R)T\sum_{n}\frac{(1-e^{-i\omega_{n}\tau})}{\omega_{n}^{2}+\omega_{r}^{2}}\nonumber \\
 & =3\frac{e^{2}/R}{2\omega_{r}}\coth(\omega_{r}/2T)(1-e^{-\omega_{r}|\tau|}).\label{bareprop}\end{align}
 Thus at long times, $\omega_{r}|\tau|\gg1,$ the correction to the
tunneling term of the AES model due to dipole modes is smaller than
the bare value by a factor of $(e^{2}/R)/\omega_{r}.$ In most common
cases of granular metals, this ratio is of the order of $10^{-2},$
as $E_{c}\sim10^{2}K$ and $\omega_{r}\sim10^{4}K.$ At short times,
$\omega_{r}|\tau|\ll1,$ the correction is smaller than the bare value
by a factor $\omega_{r}|\tau|\ll1.$ Thus under most common physical
circumstances, our approximation is valid.

For finite tunneling, the propagator for the dipole modes is that
of a damped harmonic oscillator, \begin{align}
{\mathcal{D}}_{\mathbf{ij}}^{\alpha\beta}(i\omega_{n})=\frac{\delta^{\alpha\beta}\delta_{\mathbf{i}\mathbf{j}}(e^{2}/R)}{\omega_{n}^{2}+\omega_{r}^{2}+\Gamma(\omega_{n})|\omega_{n}|},\label{D}\end{align}
 where the resonance linewidth is

\begin{align}
\Gamma & =\sum_{\mathbf{a}}\Gamma_{\mathbf{i},\mathbf{i}+\mathbf{a}}.\label{gammadef}\end{align}

\section{Optical conductivity \label{sec:Optical-conductivity}}

In this section we calculate the optical conductivity of isolated
metallic grains and then generalize it to finite intergrain tunneling.
For isolated grains, we show that the optical conductivity may be
obtained in two ways: directly from the Kubo formula, and from the
dielectric function.

\subsection{Isolated grains \label{sub:kuboIsolated}}

\subsubsection{Kubo formula approach }

We first calculate the optical conductivity for isolated grains using
the Kubo formula approach. For this we introduce an infinitesimal
vector potential $\mathbf{A}_{\tau\mathbf{x}}$ that couples to the
current $\mathbf{j}_{\mathbf{\tau\mathbf{x}}}$ and is related to
the electric field through $\mathbf{E}_{\tau\mathbf{x}}=i\partial_{\tau}\mathbf{A}_{\tau\mathbf{x}}.$
The electronic kinetic energy becomes \begin{align}
\epsilon(\tilde{\mathbf{p}}_{\tau\mathbf{x}_{\mathbf{i}}}) & \rightarrow\epsilon(\tilde{\mathbf{p}}_{\mathbf{x}_{\mathbf{i}}}-\tfrac{e}{c}\mathbf{A}_{\tau\mathbf{x}_{\mathbf{i}}})\nonumber \\
 & =\epsilon(\tilde{\mathbf{p}}_{\mathbf{x}_{\mathbf{i}}})-\tfrac{e}{mc}\mathbf{A}_{\tau\mathbf{x}_{\mathbf{i}}}\cdot\tilde{\mathbf{p}}_{\mathbf{x}_{\mathbf{i}}}+\tfrac{e^{2}}{2mc^{2}}\mathbf{A}_{\tau\mathbf{x}_{\mathbf{i}}}^{2},\label{finiteA1}\end{align}
 where $\mathbf{p}_{\mathbf{x}_{\mathbf{i}}}=-i\pmb\nabla_{\mathbf{x}_{\mathbf{i}}}$
and $\tilde{\mathbf{p}}_{\tau\mathbf{x}_{\mathbf{i}}}=\mathbf{p}_{\mathbf{x}_{\mathbf{i}}}-\frac{1}{R}\pmb\varphi_{\mathbf{i}1}(\tau)$,
and we have chosen the gauge $\nabla\cdot\mathbf{A}=0.$ The optical
conductivity tensor $\sigma^{\alpha\beta}$ is the coefficient relating
the $\mathbf{q}=0$ component of the current, \begin{align}
j_{\tau}^{\alpha}[\mathbf{A};\mathbf{q}] & =-\frac{c}{Z_{0}}\int d\mathbf{x}\, e^{-i\mathbf{q}\cdot\mathbf{x}}\int D(\text{fields})\frac{\delta S[\mathbf{A}]}{\delta A_{\tau\mathbf{x}}^{\alpha}}e^{-S[\mathbf{A}]},\label{current1}\end{align}
 to the $\mathbf{q}=0$ component of the electric field $E_{\tau\mathbf{x}}^{\beta}=i\partial_{\tau}A_{\tau\mathbf{x}}^{\beta},$\begin{align}
\sigma^{\alpha\beta}(\tau,\tau') & =Tc^{2}\sum_{m}\sigma^{\alpha\beta}(i\omega_{m})e^{-i\omega_{m}(\tau'-\tau)}\nonumber \\
 & =\frac{\delta j_{\tau}^{\alpha}[\mathbf{A};\mathbf{q}=0]}{\delta E_{\tau'}^{\beta}(\mathbf{q}=0)}\bigg|_{\mathbf{A}=0}.\label{conductivity1}\end{align}
 Here $Z_{0}=Z[\mathbf{A}=0]$. Analytically continuing to real frequencies
gives the well-known Kubo formula for the optical conductivity,\begin{align}
\sigma^{\alpha\beta}(\omega,T) & =-\frac{ic^{2}}{\omega Z_{0}{\mathcal{V}}}\int d\mathbf{x}\, d\mathbf{x}'\int_{\tau}e^{i\omega_{m}\tau}\nonumber \\
\times & \int D(\text{fields})\left[\frac{\delta^{2}S[\mathbf{A}]}{\delta A_{\tau\mathbf{x}'}^{\beta}\delta A_{0\mathbf{x}}^{\alpha}}-\frac{\delta S[\mathbf{A}]}{\delta A_{\tau\mathbf{x}'}^{\beta}}\frac{\delta S[\mathbf{A}]}{\delta A_{0\mathbf{x}}^{\alpha}}\right]\bigg|_{i\omega_{m}\rightarrow\omega}.\label{conductivity2}\end{align}
 where ${\mathcal{V}}$ is the volume of the system. The first term
in Eq.(\ref{conductivity2}), as we shall see below, represents the
inertial response of the electrons in the bulk metal. The second term,
which vanishes in the bulk, makes a finite contribution in the granular
metal. We denote these two contributions as \begin{align}
\sigma^{\alpha\beta}(\omega,T) & =\sigma_{\text{inertial}}^{\alpha\beta}(\omega,T)+\sigma_{\text{finite R}}^{\alpha\beta}(\omega,T).\label{conductivity3}\end{align}
 We can show that the {}``inertial'' term is \begin{align}
\sigma_{\text{inertial}}^{\alpha\beta}(\omega,T) & =-\frac{ie^{2}\delta^{\alpha\beta}}{\omega m{\mathcal{V}}}\sum_{\mathbf{i}}\int d\mathbf{x}_{\mathbf{i}}\langle G_{\pmb\varphi_{\mathbf{i}1}}(\mathbf{x}_{\mathbf{i}},\mathbf{x}_{\mathbf{i}};\tau,\tau)\rangle-\nonumber \\
 & -\frac{ie^{2}}{\omega m^{2}{\mathcal{V}}}\sum_{\mathbf{i}}\int d\mathbf{x}_{\mathbf{i}}d\mathbf{x}'_{\mathbf{i}}\int_{\tau}\, e^{i\omega_{m}\tau}\langle\tilde{p}_{\tau\mathbf{x}'_{\mathbf{i}}}^{\beta}\tilde{p}_{0\mathbf{x}_{\mathbf{i}}}^{\alpha}\times\nonumber \\
 & \times G_{\pmb\varphi_{\mathbf{i}1}}(\mathbf{x}'_{\mathbf{i}},\mathbf{x}_{\mathbf{i}};\tau,0)G_{\pmb\varphi_{\mathbf{i}1}}(\mathbf{x}_{\mathbf{i}},\mathbf{x}'_{\mathbf{i}};0,\tau)\rangle\bigg|_{i\omega_{m}\rightarrow\omega}.\label{diamag1}\end{align}
 This can be expressed in terms of the response function $K_{\mathbf{i}}^{\alpha\beta}$
we defined in Eq.(\ref{emr1}). For simplicity we assume that all
grains in the system are identical. Also, as in Appendix \ref{sec:app1},
we approximate the Green functions $G_{\pmb\varphi_{\mathbf{i}1}}(\mathbf{x}'_{\mathbf{i}},\mathbf{x}_{\mathbf{i}};\tau,0)$
etc. by their bare values. Then \begin{align}
\sigma_{\text{inertial}}^{\alpha\beta}(\omega,T) & =\bigg[-\frac{if}{\omega}K_{\mathbf{i}}^{\alpha\beta}(i\omega_{m})-\frac{ie^{2}}{\omega m^{2}R^{2}{\mathcal{V}}}\sum_{\mathbf{i}\lambda}\!\!\int_{\tau}e^{i\omega_{m}\tau}\times\nonumber \\
 & \times G_{\mathbf{i}\lambda}^{(0)}(\tau,0)G_{\mathbf{i}\lambda}^{(0)}(0,\tau)\langle\varphi_{\mathbf{i}1}^{\beta}(\tau)\varphi_{\mathbf{i}1}^{\alpha}(0)\rangle\bigg]\bigg|_{i\omega_{m}\rightarrow\omega}.\label{diamag2}\end{align}
 where $f$ is the volume fraction occupied by the metallic spheres.

The second term on the right hand side of Eq.(\ref{diamag2}) is smaller
than the first by a factor of $\nu(\epsilon_{F})(e^{2}/R)/(N\omega_{r}mR^{2}),$
where $N$ is the number of conduction electrons in a grain. Since
$1/(mR^{2})\sim\delta\sim\epsilon_{F}/N,$ and $\nu(\epsilon_{F})\sim N/\epsilon_{F},$
the second term is smaller by a factor of about $1/N.$ This is a
small number since the number of conduction electrons in a grain in
typical systems is of the order of $10^{4}.$ We have, dropping this
term from Eq.(\ref{diamag2}), \begin{align}
\sigma_{\text{inertial}}^{\alpha\beta}(\omega,T) & \approx-\frac{if}{\omega}K_{\mathbf{i}}^{\alpha\beta}(i\omega_{m})\bigg|_{i\omega_{m}\rightarrow\omega}\nonumber \\
 & =-\frac{ifne^{2}}{m\omega}\delta^{\alpha\beta},\label{diamag3}\end{align}
 where we used Eq.(\ref{emr4}) in the second line. This is indeed
of the form of an inductive contribution.

Now consider the finite size contribution to the conductivity described
in Eq.(\ref{conductivity2}) and Eq.(\ref{conductivity3}),\begin{align}
\sigma_{\text{finite R}}^{\alpha\beta}(\omega,T) & =\frac{ie^{2}}{\omega m^{2}{\mathcal{V}}}\sum_{\mathbf{ij}}\int d\mathbf{x}_{\mathbf{i}}d\mathbf{x}_{\mathbf{j}}\int_{\tau}\, e^{i\omega_{m}\tau}\times\nonumber \\
 & \times\langle\tilde{p}_{\tau\mathbf{x}_{\mathbf{i}}}^{\beta}\tilde{p}_{0\mathbf{x}_{\mathbf{j}}}^{\alpha}G_{\pmb\varphi_{\mathbf{i}1}}(\mathbf{x}_{\mathbf{i}},\mathbf{x}_{\mathbf{i}};\tau,\tau)\times\nonumber \\
 & \times G_{\pmb\varphi_{\mathbf{j}1}}(\mathbf{x}_{\mathbf{j}},\mathbf{x}_{\mathbf{j}};0,0)\rangle\bigg|_{i\omega_{m}\rightarrow\omega}.\label{paramag1}\end{align}
 The diagonal matrix elements of the momenta $\mathbf{p}$ are identically
zero in a finite system, $\langle\lambda|\mathbf{p}|\lambda\rangle=0$,
therefore we discard in Eq.(\ref{paramag1}) terms of the type \begin{align*}
\int d\mathbf{x}_{\mathbf{i}} & p_{\mathbf{x}_{\mathbf{i}}}^{\beta}G_{\pmb\varphi_{\mathbf{i}1}}(\mathbf{x}_{\mathbf{i}},\mathbf{x}_{\mathbf{i}};\tau,\tau)\equiv0;\end{align*}
 this simplifies the finite size contribution to \begin{align}
\sigma_{\text{finite R}}^{\alpha\beta}(\omega,T) & =\frac{ie^{2}}{\omega m^{2}R^{2}{\mathcal{V}}}\sum_{\mathbf{ij};\lambda\lambda'}\int_{\tau}\, e^{i\omega_{m}\tau}\times\nonumber \\
\times & G_{\mathbf{i}\lambda}^{(0)}(\tau,\tau)G_{\mathbf{j}\lambda'}^{(0)}(0,0)\langle\varphi_{\mathbf{i}1}^{\beta}(\tau)\varphi_{\mathbf{j}1}^{\alpha}(0)\rangle\bigg|_{i\omega_{m}\rightarrow\omega}.\label{paramag2}\end{align}
 Here we have as usual approximated the Green functions $G_{\pmb\varphi_{\mathbf{i}1}}(\mathbf{x}_{\mathbf{i}},\mathbf{x}_{\mathbf{i}};\tau,\tau)$
by the bare values. Evaluating Eq.(\ref{paramag2}) gives the following
finite size contribution, \begin{align}
\sigma_{\text{finite R}}^{\alpha\beta}(\omega,T) & =\frac{ine^{2}Nf}{\omega m^{2}R^{2}}\frac{(e^{2}/R)}{\omega_{m}^{2}+\omega_{r}^{2}}\bigg|_{i\omega_{m}\rightarrow\omega}\nonumber \\
 & =\frac{ine^{2}f}{\omega m}~\frac{\omega_{r}^{2}}{-(\omega+i0^{+})^{2}+\omega_{r}^{2}}.\label{paramag3}\end{align}
 Here $N=(4\pi/3)nR^{3}$ is the total number of conduction electrons
on a grain, $n$ is the conduction electron density, and we used $\omega_{r}^{2}=(4\pi/3)ne^{2}/m.$
Adding the inertial and finite size contributions from Eq.(\ref{diamag3})
and Eq.(\ref{paramag3}), we arrive at the optical conductivity for
isolated spherical grains, \begin{align}
\sigma^{\alpha\beta}(\omega,T) & =-\frac{ine^{2}f}{m}\frac{\delta^{\alpha\beta}\omega}{\omega^{2}-\omega_{r}^{2}-i|\omega|0^{+}}.\label{conductivity4}\end{align}
 Eq.(\ref{conductivity4}) agrees with the expression for the optical
conductivity in Eq.(\ref{sphereOptical}) that was obtained from a
simple analysis of the equation of motion of the electrons in a clean
grain. For a finite intra-grain relaxation time, $\tau_{\text{grain}},$
the optical conductivity takes the form \begin{align}
\sigma^{\alpha\beta}(\omega,T) & =-\frac{ine^{2}f}{m}\frac{\delta^{\alpha\beta}\omega}{\omega^{2}-\omega_{r}^{2}-i|\omega|/\tau_{\text{grain}}}.\label{conductivity5}\end{align}

\subsubsection{Conductivity from the dielectric function}

The optical conductivity of isolated grains that we obtained from
a tedious Kubo approach could also be inferred from the dielectric
function. In a gaussian theory, the dielectric function $\varepsilon$
\emph{for a single grain} can be extracted from the effective action,
\begin{align}
S & =-\frac{R^{3}}{2e^{2}}\sum_{\mathbf{i},l,\alpha\beta}\int_{\tau,\tau'}\varepsilon_{\mathbf{i}l}^{\alpha\beta}(\tau-\tau')E_{\mathbf{i}l}^{\alpha}(\tau)E_{\mathbf{i}l}^{\beta}(\tau'),\label{gaussian}\end{align}
 where $E_{\mathbf{i}l}^{\alpha}(\tau)$ are the multipole components
of the electric field (see Eq.(\ref{hubbard2})ff) at the grains.
The lowest possible angular momentum component of an excitation on
an isolated grain is $l=1$. That is, in the absence of intergrain
tunneling, the simplest response to an electric field is a uniform
polarization. Thus we need to consider only $E_{\mathbf{i}1}^{\alpha}(\tau)=V_{\mathbf{i}1}^{\alpha}/R=i\partial_{\tau}\varphi_{\mathbf{i}1}^{\alpha}(\tau)/R.$
Furthermore, because of the high energy $\omega_{r}\sim e\sqrt{n/m}$
associated with the dipole excitations, we can safely neglect in the
effective action terms with higher powers of $\pmb\varphi_{\mathbf{i}1}(\tau).$

The following relation can be gathered from Eq.(\ref{seffbare}),
Eq.(\ref{conductivity4}) and Eq.(\ref{gaussian}), \begin{align}
\sigma^{\alpha\beta}(\omega,T)\bigg|_{g=0} & \!\!\!\!\!\!=-\frac{ine^{2}f}{m\omega}\frac{\delta^{\alpha\beta}}{\varepsilon_{\mathbf{i}1}(\omega,T)}=-\frac{ine^{2}f}{m}\frac{\delta^{\alpha\beta}\omega}{\omega^{2}-\omega_{r}^{2}-i\eta|\omega|}.\label{dielfn1}\end{align}
 Such a cross relation has been discussed, for instance, by Hopfield\cite{hopfield1965}
in 1965. Physically, the imaginary part of the dielectric function
is associated with relaxation, so a stronger relaxation implies weaker
conduction.

\subsection{Finite intergrain tunneling}

The Kubo approach is the most reliable way to calculate the optical
conductivity, but, as illustrated in Sec.\ref{sub:kuboIsolated},
it is very tedious even for an isolated sphere. At finite intergrain
tunneling, an even larger number of terms involving both intragrain
and intergrain currents would have to be calculated. We also showed
that for gaussian models, the dielectric function could be used to
obtain the conductivity with significantly less effort. However, as
the discussion below shows, the theory is gaussian only in the two
extreme cases of isolated grains, $g=0,$ or strongly coupled grains,
$g\gg1.$ So we resort to a combination of the Kubo and dielectric
function approach, using the Kubo approach for multipole modes that
cannot be considered in a gaussian approximation, and retaining the
dielectric function approach for modes that are effectively gaussian.

At finite intergrain tunneling, an electric field can cause intergrain
polarization (opposite charges on adjacent grains) as well as intergrain
polarization. We must therefore consider the contribution of the monopole
modes $\varphi_{\mathbf{i}0}(\tau)$ in the dielectric response function.
Tunneling events are accompanied by fluctuations in electrostatic
energy which can be large, of the order of $e^{2}/R,$ when intergrain
tunneling is weak, $g\ll1.$ Therefore for weak but finite tunneling,
we must consider non-gaussian contributions for the monopole modes,
$V_{\mathbf{i}0}(\tau)=i\partial_{\tau}\varphi_{\mathbf{i}0}(\tau).$
This is clear from the effective field theory at finite tunneling
given by Eq.(\ref{Spol}) and Eq.(\ref{SAES}). On the other hand,
for strong intergrain tunneling, $g\gg1,$ monopole fluctuations are
small because charges can easily flow to neutralize potential differences
between the grains. In this case, we again have an approximately gaussian
theory for the $\varphi_{\mathbf{i}0}(\tau)$ modes. We write the
total conductivity as a sum of the monopole and dipole contributions,
\begin{align}
\sigma^{\alpha\beta}(\omega,T) & \approx\sigma_{0}^{\alpha\beta}(\omega,T)+\sigma_{1}^{\alpha\beta}(\omega,T).\label{conductivity}\end{align}
 The conductivity due to the monopole part has been obtained elsewhere\cite{efetov02}
in the context of the AES model, \begin{align}
\sigma_{0}^{\alpha\beta}(\omega,T) & =\frac{ia^{2-d}}{\omega}\int_{\tau}\, e^{i\Omega_{n}\tau}K_{\mathrm{AES}}^{\alpha\beta}(\tau)\big|_{\Omega_{n}\rightarrow-i\omega},\label{KAES}\end{align}
 where $a$ is the intergrain distance. It consists of diamagnetic
and paramagnetic parts,\begin{align*}
K_{\mathrm{AES}}^{\alpha\beta}(\tau) & =K_{\mathrm{AES}}^{\alpha\beta,dia}(\tau)+K_{\mathrm{AES}}^{\alpha\beta,para}(\tau),\end{align*}
 \begin{align}
K_{\mathrm{AES}}^{\alpha\beta,dia}(\tau) & =\delta^{\alpha\beta}e^{2}\pi gT^{2}\int_{\tau'}(\delta(\tau)-\delta(\tau'-\tau))\times\nonumber \\
\times & \frac{1}{\sin^{2}(\pi T\tau')}\langle\cos(\varphi_{\mathbf{i},\mathbf{i}+\mathbf{e}_{\alpha},0}(\tau)-\varphi_{\mathbf{i},\mathbf{i}+\mathbf{e}_{\alpha},0}(\tau'))\rangle,\label{KAESdia}\end{align}
 \begin{align}
K_{\mathrm{AES}}^{\alpha\beta,para}(\tau) & =-\delta^{\alpha\beta}\sum_{\mathbf{i}}\langle X_{0}^{\alpha}(\tau)X_{\mathbf{i}}^{\alpha}(0)\rangle,\label{KAESpara}\end{align}
 where \begin{align}
X_{\mathbf{i}}^{\gamma}(\tau) & =e\pi gT^{2}\int_{\tau'}\frac{1}{\sin^{2}[\pi T(\tau-\tau')]}\times\nonumber \\
 & \times\langle\sin(\varphi_{\mathbf{i},\mathbf{i}+\mathbf{e}_{\gamma},0}(\tau)-\varphi_{\mathbf{i},\mathbf{i}+\mathbf{e}_{\gamma},0}(\tau'))\rangle.\label{AEScurrent}\end{align}
 In order to remind us of the AES origin of the $l=0$ component of
the conductivity, we rename $\sigma_{0}(\omega,T)$ to $\sigma_{\mathrm{AES}}(\omega,T).$

The contribution to the conductivity from the dipole part is written
in terms of the $l=1$ component of the dielectric function (see discussion
above and in the previous section), \begin{align}
\sigma_{1}^{\alpha\beta}(\omega,T) & =-\frac{ine^{2}f}{m\omega}\frac{\delta^{\alpha\beta}}{\varepsilon_{\mathbf{i}1}(\omega,T)}\nonumber \\
 & =-\frac{ine^{2}f}{m}\frac{\delta^{\alpha\beta}\omega}{\omega^{2}-\omega_{r}^{2}-i\Gamma(\omega,T)|\omega|}.\label{dipolecond}\end{align}
 Here $\Gamma(\omega,T)$ is the intergrain relaxation rate defined
in Eq.(\ref{Gamma1}). $\Gamma$ involves the \emph{cosine} correlator
$\Pi$ defined in Eq.(\ref{ctaudef}), and thus closely resembles
the \emph{diamagnetic} part of the AES conductivity, Eq.(\ref{KAESdia}).
However, the full AES conductivity behaves very differently because
of the \emph{paramagnetic} contribution, Eq.(\ref{KAESpara}).

In the presence of intragrain relaxation mechanisms such as impurity
scattering or boundary scattering, we expect that \begin{align}
\sigma^{\alpha\beta}(\omega,T) & =\delta^{\alpha\beta}\sigma_{\mathrm{AES}}(\omega,T)-\frac{ine^{2}f}{m}\frac{\delta^{\alpha\beta}\omega}{\omega^{2}-\omega_{r}^{2}-i|\omega|/\tau_{\text{rel}}}\label{final-opticalcond}\end{align}
 for consistency with Eq.(\ref{sphereOptical2}) and Eq.(\ref{conductivity5}),
where the total relaxation rate $\tau_{\text{rel}}^{-1}$ is given
by the Matthiessen rule, $\tau_{\text{rel}}^{-1}=\tau_{\text{grain}}^{-1}+\Gamma.$

\subsection{Some special cases}

The final expression for the optical conductivity contains the AES
conductivity $\sigma_{\mathrm{AES}}(\omega,T)$ and the resonance
width due to intergrain tunneling $\Gamma(\omega,T).$ $\sigma_{\mathrm{AES}}(\omega,T)=\sigma_{0}(\omega,T)$
has been explicitly defined in Eq.(\ref{KAES}) through Eq.(\ref{AEScurrent}),
and $\Gamma(\omega,T)$ has been defined in Eq.(\ref{Gamma1}) and
Eq.(\ref{gammadef}). Below we discuss a few special cases for a regular
three dimensional array. Throughout we assume that the frequency lies
in the range $\omega/T\gg1,\,\omega\tau_{c}\sim\omega/\epsilon_{F}\ll1,$
and the temperature much smaller than the charging energy, $E_{c}/T\gg1.$

\textbf{(a)} Consider first small intergrain tunneling conductance,
$g\ll1.$ As the calculations are very complicated, we refer the reader
to Appendices \ref{sec:AEScond} and \ref{sec:reswidth} for the details.
We show there that at frequencies much larger than the charging energy,
the conductivity tends to saturate, $\sigma_{\mathrm{AES}}\approx g(e^{2}/a).$
The same goes for the polarization resonance width, $\Gamma\approx4bzg(e^{2}/R),$
where $z$ is the grain coordination number and we used Eq.(\ref{Gamma1})
and Eq.(\ref{gammadef}). If the frequency is much smaller than the
charging energy, the conductivity is dominated by thermal excitation
of quasiparticles and obeys an Arrhenius law,\begin{align}
\sigma_{\mathrm{AES}} & \approx2g\frac{e^{2}}{a}e^{-E_{c}/T},\,\,\,\omega\ll E_{c}.\label{lowfreqAES}\end{align}
 In contrast, the resonance width does not obey an Arrhenius law:\begin{align}
\Gamma & \approx4bzg\frac{e^{2}}{R}\frac{4\pi}{3E_{c}^{2}}[T^{2}+(\omega/2\pi)^{2}],\,\,\,\omega\ll E_{c}.\label{lowfreqGamma}\end{align}
 Suppose the charging energy is small compared to the resonance frequency,
$E_{c}\ll\omega_{r}\ll\epsilon_{F}.$ As significant changes in $\sigma_{\mathrm{AES}}$
and $\Gamma$ occur on the scale of the Coulomb blockade energy $E_{c},$
the frequencies in the vicinity of the resonance are too large for
Coulomb blockade physics to be significant. In this case, $\Gamma\approx4bzg^{2}(e^{2}/R)$
is practically independent of frequency and temperature.

Consider now the case where charging energy is large or comparable
with respect to the resonance frequency, $\omega_{r}<E_{c}\ll\epsilon_{F}.$
This can happen if the metal has a low enough conduction electron
density, a large effective mass for the electrons, and/or small grains.
Increasing the volume fraction of the metal is another way in which
the resonance frequency may be reduced; we shall see in Sec.\ref{Drudecomparison}
that $\omega_{r}$ renormalizes to $\omega_{r}^{*}=\omega_{r}\sqrt{1-f}$
as $f$ is increased. This regime is very interesting because the
resonance is in the low frequency regime $(\omega\ll E_{c})$ for
Coulomb-blockade physics. So near the resonance $\omega=\omega_{r}^{*}$,
while $\sigma_{\mathrm{AES}}$ still obeys an Arrhenius law, Eq.(\ref{lowfreqAES}),
the temperature dependence of $\Gamma$ can be qualitatively different
from $\sigma_{\mathrm{AES}}.$ One expects here, following Eq.(\ref{lowfreqGamma}),
$\Gamma(\omega_{r}^{*},T)\propto(T^{2}+(\frac{\omega_{r}^{*}}{2\pi})^{2})/E_{c}^{2}.$

\textbf{(b)} Finally consider large intergrain conductance, $g\gg1.$
In this case both $\sigma_{\mathrm{AES}}(\omega,T)$ and $\Gamma(\omega,T)$
evolve logarithmically\cite{efetov02} with temperature and frequency,
\begin{align*}
\sigma_{\mathrm{AES}}(\omega,T) & \approx g(e^{2}/a)\left[1-\frac{1}{\pi gz}\ln\left(\frac{gE_{c}}{\text{max}(\omega,T)}\right)\right],\\
\Gamma(\omega,T) & \approx4bzg\frac{e^{2}}{R}\left[1-\frac{1}{\pi gz}\ln\left(\frac{gE_{c}}{\text{max}(\omega,T)}\right)\right],\end{align*}
 down to exponentially low temperatures and frequencies when perturbation
theory is no longer valid. Below such low temperatures, the physics
is similar to the $g\ll1$ case discussed above.

\subsection{Comparison with Drude theory}

\label{Drudecomparison} According to Drude theory, a bulk metal will
have a frequency-dependent dielectric function \begin{equation}
\varepsilon_{\text{Drude}}(\omega)=1-\frac{4\pi ne^{2}}{m}\frac{1}{\omega(\omega+i/\tau_{\text{Drude}})},\label{epsilonDrude}\end{equation}
 which when substituted in the Maxwell-Garnett formula, Eq.(\ref{maxwellgarnett}),
yields the effective dielectric function for a homogeneous system
of metallic grains in vacuum, \begin{equation}
\varepsilon_{\text{eff}}(\omega)=\frac{\omega^{2}-(4\pi ne^{2}/3m)(1+2f)-i\omega/3\tau_{\text{Drude}}}{\omega^{2}-(4\pi ne^{2}/3m)(1-f)-i\omega/3\tau_{\text{Drude}}}.\label{epsilonMG}\end{equation}
 From Eq.(\ref{epsilonMG}) one then infers the optical conductivity
$\sigma_{MG}(\omega)$ for the granular system in the Maxwell-Garnett
approximation, \begin{equation}
\text{Re }\sigma_{MG}(\omega)=\frac{fne^{2}}{m}\frac{\omega^{2}/(3\tau_{\text{Drude}})}{[\omega^{2}-\omega_{r}^{2}(1-f)]^{2}+(\omega/3\tau_{\text{Drude}})^{2}}.\label{sigmaMG}\end{equation}
 Eq.(\ref{sigmaMG}) is not strictly correct because $\tau_{\text{Drude}}$
does not include the Landau damping\cite{kawabata1966,wood1982,kreibig1985}
that exists in the metallic grain but is absent in the bulk. (See
also the discussion in Sec.\ref{sec:Effective-field}.) Besides, matching
Eq.(\ref{sigmaMG}) to the correct DC conductivity requires that $\tau_{\text{Drude}}\sim\sigma_{\text{AES}}(T)\sim\exp(-E_{c}/T)$,
whereas we have shown that the resonance width $\Gamma$ has a \emph{different}
temperature dependence from $\sigma_{\text{AES}}$. Evidently, classical
arguments are unable to explain the full behavior of $\sigma(\omega,T)$.

Pending a proper theory of long-range interaction of dipoles in the
granular metal, we nevertheless propose that the optical conductivity
of the granular metal in the Maxwell-Garnett approximation is given
by Eq.(\ref{sigmaMG}) with $\tau_{\text{rel}}^{-1}=\tau_{\text{grain}}^{-1}+\Gamma$
replacing $(3\tau_{\text{Drude}})^{-1}$: \begin{align}
\text{Re }\sigma(\omega,T) & =\sigma_{\text{AES}}(\omega,T)+\nonumber \\
 & +\frac{fne^{2}}{m}\frac{\omega^{2}/\tau_{\text{rel}}}{[\omega^{2}-\omega_{r}^{2}(1-f)]^{2}+(\omega/\tau_{\text{rel}})^{2}}.\label{sigmaMG2}\end{align}
 The Maxwell-Garnett result, Eq.(\ref{sigmaMG2}), agrees with the
dipole contribution in our stronger result for the optical conductivity,
Eq.(\ref{final-opticalcond}), that was derived for a dilute granular
array, $f\ll1$. Eq.(\ref{sigmaMG2}) shows that the resonance frequency
undergoes an infrared shift, $\omega_{r}^{*}=\omega_{r}\sqrt{1-f}$,
as the volume fraction of the metal is increased. This dependence
has been previously obtained\cite{stroud1979} in the literature.
One must take care not to extend Eq.(\ref{sigmaMG2}) all the way
to $f=1$ because the validity of our effective field theory is limited
to the insulating phase of the granular metal.

\section{Conclusions and discussion \label{sec:Conclusions}}

We have developed an effective field theory of granular metals which
is a generalization of the Ambegaokar-Eckern-Schön (AES) action to
include polarization degrees of freedom. This approach synthesizes
the classical electrodynamic theories of Maxwell Garnett and Mie and
the quantum mechanical AES model for dissipative transport in order
to capture both finite-frequency and finite-temperature effects. It
is valid at temperatures larger than the mean level spacing $\delta$
in a grain.

Using this effective field theory, we have calculated the frequency
and temperature dependence of the optical conductivity of an array
of spherical metallic grains. We have shown that the temperature dependence
of the polarization resonance width $\Gamma$ differs qualitatively
from that of the DC conductivity for frequencies and temperatures
much smaller than the charging energy of the grains. While the DC
conductivity obeys an Arrhenius law at low temperatures, $\Gamma$
decreases only algebraically as a function of frequency and temperature.
We believe this prediction can be tested in experimental situations
where the condition $\omega_{r}\sqrt{1-f}<E_{c}$ can be satisfied.
This can occur in systems where the conduction electron density is
low, the effective mass is large, and/or the grains are small, and
the volume fraction of the metal is large (while still remaining in
the insulating phase). This qualitative difference between the temperature
dependences of the DC conductivity $\sigma(0,T),$ and the collective
mode damping $\Gamma(0,T),$ obeyed in certain granular metals is
quite unlike the behavior seen\cite{blumberg2002} in pinned sliding
density wave compounds where the temperature dependence of the collective
mode damping is the same as the DC conductivity. Such a difference
could perhaps be used to distinguish between granularity arising from
spontaneous electronic phase segregation in strongly correlated electron
systems and density wave order.

To keep our analysis simple, we have, in our field theoretical treatment,
ignored electrostatic interactions between monopoles (charges) and
dipoles (polarizations) on different grains. Strictly speaking, this
is correct only in a dilute granular array $(f\ll1)$ or at frequencies
higher than the polarization resonance. Renormalization of the resonance
frequency due to the presence of neighboring grains, even in the absence
of tunneling, is one effect that is lost in this approximation. Pending
a general field theoretical treatment of long-range interaction of
dipoles, we have used our result for the optical response of a dilute
array of grains as an input in a Maxwell-Garnett effective medium
approximation to obtain the optical conductivity at larger values
of $f$. The shift that we obtain in the resonance frequency as a
function of $f$ agrees with earlier results in the literature.\cite{stroud1979}

Another aspect we have not considered is disorder, both in intergrain
tunneling conductance and as a random background potential due to
quenched impurities in the insulating part. In presence of strong
disorder, the DC conductivity obeys a soft-activation law $\sigma(0,T)\sim\sigma_{0}e^{-\sqrt{T_{0}/T}}$
instead of an Arrhenius law; it should be interesting to consider
the effect on optical conductivity. In principle it is possible to
study the effect of both kinds of disorder in our scheme.

Finally there are some fundamental limitations on AES-inspired treatments.
Like the AES model, our dissipative transport model is limited to
the insulating side of a metal-insulator transition, and cannot describe
the optical conductivity through the transition; it also neglects
quantum coherence effects, which are important at $T<\delta$.

The present level of rigor in our calculation is insufficient to study
the various $f-$sum rules obeyed by the optical conductivity.\cite{stroud1979}
Our model is justified only for frequencies much smaller than the
bandwidth, $\omega\ll\tau_{c}^{-1}\sim\epsilon_{F}.$ At higher frequencies,
or in other words for times shorter than the cutoff, $\tau\ll\tau_{c}$
(see discussion following Eq.(\ref{localgreenfn})), the dissipation
kernel, $T^{2}/\sin^{2}(\pi T\tau),$ that appears in the tunneling
terms in the effective field theory, Eq.(\ref{stunn3}), is no longer
valid.

\begin{acknowledgments}
We are grateful to G.~Blumberg, D.~E.~Khmelnitskii, \v{S}imon
Kos, and P.~B.~Littlewood for valuable discussions. VT thanks Trinity
College, Cambridge for a JRF. YLL thanks Purdue University for support. 
\end{acknowledgments}
\appendix

\section{Effect of long-range Coulomb interaction on optical conductivity
of dirty insulators \label{sec:app0}}

Shklovskii and Efros generalized Mott's treatment to include the effect
of long-range Coulomb interactions.\cite{efros1981,shklovskii81}
The difference is particularly significant when the particle-hole
Coulomb energy at hopping distance $r_{\omega}$ exceeds the optical
frequency: $e^{2}/\varepsilon r_{\omega}\gg\omega.$ Here $\varepsilon$
is the dielectric constant of the medium. Physically, in the presence
of Coulomb interactions, transitions to the final state $\epsilon_{j}$
can be made from an occupied level with energy in the range $\epsilon_{F}-\omega-\frac{e^{2}}{\varepsilon r_{\omega}}<\epsilon_{i}\leq\epsilon_{F}.$
Modifying the limits in Eq.(\ref{mott1}) accordingly,\cite{shklovskii81}\begin{align}
\text{Re}(\sigma(\omega)) & \sim2\pi e^{2}n_{\text{imp}}^{2}\frac{\omega}{\delta^{2}}\left(\omega+\frac{e^{2}}{\varepsilon r_{\omega}}\right)(r_{\omega}^{d-1}\xi_{loc})r_{\omega}^{2}\nonumber \\
 & \approx2\pi e^{4}n_{\text{imp}}^{2}\frac{\omega}{\varepsilon\delta^{2}}\xi_{loc}^{d+1}\ln^{d}(w/\omega),\, e^{2}/\varepsilon r_{\omega}\gg\omega.\label{shklosefros}\end{align}
 Eq.(\ref{shklosefros}) assumes that the density of states is a constant;
and this is correct as long as the energy $(\omega+\frac{e^{2}}{\varepsilon r_{\omega}})$
is larger than the Coulomb gap, $\Delta.$ At energies less than $\Delta,$
the density of states at the chemical potential is not a constant,
but instead has the form $\rho(\epsilon)\sim|\epsilon|^{d-1}(\varepsilon/e^{2})^{d}.$
The Coulomb gap is the energy at which the density of states reaches
the value in the absence of Coulomb interaction; thus $\Delta\sim[e^{2d}n_{\text{imp}}/(\varepsilon\delta)]^{1/(d-1)}.$
Using this density of states, we get, for $\Delta>\frac{e^{2}}{\varepsilon r_{\omega}}>\omega,$
an optical conductivity \begin{align}
\text{Re}(\sigma(\omega)) & \propto\omega r_{\omega}^{2-d}\sim\omega\xi_{loc}^{2-d}\ln^{2-d}(w/\omega).\label{shklosefros2}\end{align}
 At a finite temperature, $\sigma(\omega=0)$ is finite (see above).
If the temperature is low, $(T,\omega)\ll\frac{e^{2}}{\varepsilon r_{\omega}},$
the frequency dependent conductivity in Eq.(\ref{shklosefros}) has
an extra Boltzmann factor,\begin{align}
\text{Re}(\sigma(\omega,T)) & \sim\sigma(0,T)+2\pi e^{2}n_{\text{imp}}^{2}(1-e^{-\omega/T})\times\nonumber \\
 & (\omega/\varepsilon\delta^{2})\xi_{loc}^{d+1}\ln^{d}(w/\omega).\label{shklosefros3}\end{align}
 For high enough temperatures, $T\gg(\omega,\frac{e^{2}}{\varepsilon r_{\omega}}),$or
high enough frequencies, $\omega\gg(T,\frac{e^{2}}{\varepsilon r_{\omega}}),$
Mott's result, Eq.(\ref{mott2}) is obtained.\cite{shklovskii81}

The above treatment assumes that there is no inelastic scattering
(e.g., by phonons). At finite temperature, phonons (with characteristic
frequency $\omega_{ph}\sim10^{12}\text{Hz}$) provide an additional
relaxation mechanism. Electrons make transitions by emitting or absorbing
phonons with energy of the order of $\omega_{rel}\sim\omega_{ph}e^{-x_{ij}/\xi_{loc}},$
so the main contribution to the optical conductivity from inelastic
processes comes from frequencies of the order of $\omega_{rel}.$
For such frequencies, we should use $\omega_{ph}$ instead of $w$
in Eq.(\ref{shklosefros3}).

\section{Effective action corrections from expansion of $G_{\pmb\varphi_{\mathbf{i}1}}$
in powers of $\pmb\varphi_{\mathbf{i}1}$ \label{sec:app1}}

We explain how corrections to the effective tunneling action in Eq.(\ref{stunn1})
coming from the expansion of $G_{\pmb\varphi_{\mathbf{i}1}}(\mathbf{x}_{\mathbf{i}},\mathbf{x}_{\mathbf{i}};\tau,\tau')$
in powers of $\pmb\varphi_{\mathbf{i}1}$ are small compared to the
bare value when $T\gg\delta.$ We expand Eq.(\ref{Ginv}), \begin{align*}
G_{\pmb\varphi_{\mathbf{i}1}} & =\left(1-\frac{G_{\mathbf{i}}^{(0)}}{2mR^{2}}\pmb\varphi_{\mathbf{i}1}^{2}+\frac{G_{\mathbf{i}}^{(0)}}{mR}\pmb\varphi_{\mathbf{i}1}\cdot\mathbf{p}_{\mathbf{x}_{\mathbf{i}}}\right)^{-1}G_{\mathbf{i}}^{(0)},\end{align*}
 where \begin{align*}
G_{\mathbf{i}}^{(0)}(\mathbf{x}_{\mathbf{i}},\mathbf{x}_{\mathbf{i}}';\tau,\tau') & =T\sum_{\lambda,n}\frac{\psi_{\lambda}(\mathbf{x}_{\mathbf{i}}')\psi_{\lambda}^{*}(\mathbf{x}_{\mathbf{i}})}{i\nu_{n}-\xi_{\mathbf{i}\lambda}}e^{-i\nu_{n}(\tau-\tau')}\\
 & =\sum_{l}G_{\mathbf{i}\lambda}^{(0)}(\tau,\tau')\psi_{\lambda}(\mathbf{x}_{\mathbf{i}}')\psi_{\lambda}^{*}(\mathbf{x}_{\mathbf{i}}),\end{align*}
 up to second order in $\pmb\varphi_{\mathbf{i}1}.$ The resulting
correction to the effective action of Eq.(\ref{stunn2}) is\begin{align}
\delta S_{\mathrm{eff}}^{tun}[\pmb\varphi_{\mathbf{i}1}]=\frac{|t|^{2}}{4mR^{2}}\sum_{\mathbf{ia};\lambda_{1}\lambda_{2}}\int_{\tau\tau'}\int_{\tau_{1}}G_{\mathbf{i}\lambda_{1}}^{(0)}(\tau,\tau_{1})\times\nonumber \\
\times G_{\mathbf{i}\lambda_{1}}^{(0)}(\tau_{1},\tau')G_{\mathbf{i}+\mathbf{a},\lambda_{2}}^{(0)}(\tau',\tau)\langle\Pi_{\mathbf{i},\mathbf{i}+\mathbf{a}}(\tau,\tau')\rangle\pmb\varphi_{\mathbf{i}1}^{2}(\tau_{1})+\nonumber \\
+\frac{|t|^{2}}{2m^{2}R^{2}}\sum_{\mathbf{ia};\lambda_{1}\lambda_{2}\lambda_{3}}\int_{\tau\tau'}\int_{\tau_{1}\tau_{2}}|p_{\mathbf{i},\lambda_{1}\lambda_{2}}^{\alpha}|^{2}G_{\mathbf{i}\lambda_{1}}^{(0)}(\tau,\tau_{1})\times\nonumber \\
\times G_{\mathbf{i}\lambda_{2}}^{(0)}(\tau_{1},\tau_{2})G_{\mathbf{i}\lambda_{1}}(\tau_{2},\tau')G_{\mathbf{i}+\mathbf{a},\lambda_{3}}^{(0)}(\tau',\tau)\langle\Pi_{\mathbf{i},\mathbf{i}+\mathbf{a}}(\tau,\tau')\rangle\times\nonumber \\
\times\pmb\varphi_{\mathbf{i}1}(\tau_{1})\cdot\pmb\varphi_{\mathbf{i}1}(\tau_{2}),\qquad\label{stunncorr}\end{align}
 and $\Pi_{\mathbf{i},\mathbf{i}+\mathbf{a}}(\tau,\tau')$ is as defined
in Eq.(\ref{ctaudef}). Next we simplify Eq.(\ref{stunncorr}) by
integrating over $\tau,\tau'.$ Using the frequency representation
and completing the integration over $\tau,\tau'$ we have\begin{align}
\delta S_{\mathrm{eff}}^{tun}[\pmb\varphi_{\mathbf{i}1}]=\frac{|t|^{2}T^{2}}{4mR^{2}}\sum_{\mathbf{ia};\lambda_{1}\lambda_{2};n,m}\int_{\tau_{1}}\langle\Pi_{\mathbf{i},\mathbf{i}+\mathbf{a}}(\omega_{m})\rangle\pmb\varphi_{\mathbf{i}1}^{2}(\tau_{1})\times\nonumber \\
\times\frac{1}{(i\nu_{n}-\xi_{\mathbf{i}\lambda_{1}})^{2}}\frac{1}{[i(\nu_{n}+\omega_{m})-\xi_{\mathbf{i}+\mathbf{a},\lambda_{2}}]}+\nonumber \\
+\frac{|t|^{2}T^{2}}{2m^{2}R^{2}}\!\!\!\!\sum_{\mathbf{ia};\lambda_{1}\lambda_{2}\lambda_{3};n,m}\int_{\tau_{1}\tau_{2}}\!\!\!\!\! G_{\mathbf{i}\lambda_{2}}^{(0)}(\tau_{1},\tau_{2})\pmb\varphi_{\mathbf{i}1}(\tau_{1})\cdot\pmb\varphi_{\mathbf{i}1}(\tau_{2})\times\nonumber \\
\times\frac{e^{-i\nu_{n}(\tau_{2}-\tau_{1})}}{(i\nu_{n}-\xi_{\mathbf{i}\lambda_{1}})^{2}}\times\frac{|p_{\mathbf{i},\lambda_{1}\lambda_{2}}^{\alpha}|^{2}\langle\Pi_{\mathbf{i},\mathbf{i}+\mathbf{a}}(\omega_{m})\rangle}{[i(\nu_{n}+\omega_{m})-\xi_{\mathbf{i}+\mathbf{a},\lambda_{3}}]}.\qquad\label{stunncorr2}\end{align}
 It is convenient to perform the Matsubara sum over the fermionic
frequencies. We have \begin{align}
T\sum_{n}\frac{e^{-i\nu_{n}(\tau_{2}-\tau_{1})}}{(i\nu_{n}-\xi_{\mathbf{i}\lambda})^{2}}\frac{1}{[i(\nu_{n}+\omega_{m})-\xi_{\mathbf{i}+\mathbf{a},\lambda'}]}\qquad\qquad\nonumber \\
=\frac{\partial}{\partial\xi_{\mathbf{i}\lambda}}\left[\frac{G_{\mathbf{i}\lambda}^{(0)}(\tau_{2},\tau_{1})-G_{\mathbf{i}+\mathbf{a},\lambda'}^{(0)}(\tau_{2},\tau_{1})e^{i\omega_{m}(\tau_{2}-\tau_{1})}}{i\omega_{m}+\xi_{\mathbf{i}\lambda}-\xi_{\mathbf{i}+\mathbf{a},\lambda'}}\right].\label{fermisum}\end{align}
 The first term in Eq.(\ref{stunncorr2}) vanishes when we use Eq.(\ref{fermisum})
with $\tau_{1}=\tau_{2}.$ Hence \begin{align}
\delta S_{\mathrm{eff}}^{tun}[\pmb\varphi_{\mathbf{i}1}]=\frac{|t|^{2}T}{2m^{2}R^{2}}\!\!\!\!\sum_{\mathbf{ia};\lambda_{1}\lambda_{2}\lambda_{3};m}\int_{\tau_{1}\tau_{2}}\!\!\!\!\!|p_{\mathbf{i},\lambda_{1}\lambda_{2}}^{\alpha}|^{2}\times\nonumber \\
\times G_{\mathbf{i}\lambda_{2}}^{(0)}(\tau_{1},\tau_{2})\pmb\varphi_{\mathbf{i}1}(\tau_{1})\cdot\pmb\varphi_{\mathbf{i}1}(\tau_{2})\langle\Pi_{\mathbf{i},\mathbf{i}+\mathbf{a}}(\omega_{m})\rangle\times\nonumber \\
\times\frac{\partial}{\partial\xi_{\mathbf{i}\lambda_{1}}}\left[\frac{G_{\mathbf{i}\lambda_{1}}^{(0)}(\tau_{2},\tau_{1})-G_{\mathbf{i}+\mathbf{a},\lambda_{3}}^{(0)}(\tau_{2},\tau_{1})e^{i\omega_{m}(\tau_{2}-\tau_{1})}}{i\omega_{m}+\xi_{\mathbf{i}\lambda_{1}}-\xi_{\mathbf{i}+\mathbf{a},\lambda_{3}}}\right].\label{stunncorr3}\end{align}
 Eq.(\ref{stunncorr3}) contains two terms: one where the Green functions
are on the same grain and the other where they are on different grains.
The term with the Green functions on the same grain can be simplified
by summing over $\xi_{\mathbf{i}+\mathbf{a},\lambda_{3}}.$ The result
of the summation in $\nu(\epsilon_{F})\text{sgn }\omega_{m}.$ Since
$\langle\Pi_{\mathbf{i},\mathbf{i}+\mathbf{a}}(\omega_{m})\rangle$
is an even function of $\omega_{m},$ summing over $\omega_{m}$ makes
the first term disappear. Thus so far, \begin{align}
\delta S_{\mathrm{eff}}^{tun}[\pmb\varphi_{\mathbf{i}1}]=-\frac{|t|^{2}T}{2m^{2}R^{2}}\!\!\!\!\sum_{\mathbf{ia};\lambda_{1}\lambda_{2}\lambda_{3};m}\int_{\tau_{1}\tau_{2}}\!\!\!\!\! G_{\mathbf{i}\lambda_{2}}^{(0)}(\tau_{1},\tau_{2})\times\nonumber \\
\times|p_{\mathbf{i},\lambda_{1}\lambda_{2}}^{\alpha}|^{2}\pmb\varphi_{\mathbf{i}1}(\tau_{1})\cdot\pmb\varphi_{\mathbf{i}1}(\tau_{2})G_{\mathbf{i}+\mathbf{a},\lambda_{3}}^{(0)}(\tau_{2},\tau_{1})\times\nonumber \\
\times\langle\Pi_{\mathbf{i},\mathbf{i}+\mathbf{a}}(\omega_{m})\rangle\frac{\partial}{\partial\xi_{\mathbf{i}\lambda_{1}}}\frac{e^{i\omega_{m}(\tau_{2}-\tau_{1})}}{i\omega_{m}+\xi_{\mathbf{i}\lambda_{1}}-\xi_{\mathbf{i}+\mathbf{a},\lambda_{3}}}.\qquad\label{stunncorr4}\end{align}
 Now we integrate Eq.(\ref{stunncorr4}) by parts with respect to
the variable $\xi_{\mathbf{i}\lambda_{1}}$ using $\sum_{\lambda_{1}}\leftrightarrow\nu(\epsilon_{F})\int d\xi_{\mathbf{i}\lambda_{1}}$
and the identity $\partial/\partial\xi_{\mathbf{i}\lambda_{1}}=2m\sum_{\lambda}(1/p_{\mathbf{i},\lambda\lambda_{1}})(\partial/\partial p_{\mathbf{i},\lambda_{1}\lambda}),$\begin{align}
\delta S_{\mathrm{eff}}^{tun}[\pmb\varphi_{\mathbf{i}1}]=\frac{|t|^{2}T\nu(\epsilon_{F})}{mR^{2}}\!\!\!\!\sum_{\mathbf{ia};\lambda_{2}\lambda_{3};m}\int d\xi_{\mathbf{i}\lambda_{1}}\int_{\tau_{1}\tau_{2}}\!\!\!\!\! G_{\mathbf{i}\lambda_{2}}^{(0)}(\tau_{1},\tau_{2})\times\nonumber \\
\times\pmb\varphi_{\mathbf{i}1}(\tau_{1})\cdot\pmb\varphi_{\mathbf{i}1}(\tau_{2})G_{\mathbf{i}+\mathbf{a},\lambda_{3}}^{(0)}(\tau_{2},\tau_{1})\times\nonumber \\
\times\langle\Pi_{\mathbf{i},\mathbf{i}+\mathbf{a}}(\omega_{m})\rangle\frac{e^{i\omega_{m}(\tau_{2}-\tau_{1})}}{i\omega_{m}+\xi_{\mathbf{i}\lambda_{1}}-\xi_{\mathbf{i}+\mathbf{a},\lambda_{3}}}.\qquad\label{stunncorr5}\end{align}
 Now we complete the integration over $\xi_{\mathbf{i}\lambda_{1}}$
to get \begin{align}
\delta S_{\mathrm{eff}}^{tun}[\pmb\varphi_{\mathbf{i}1}]=\frac{2\pi|t|^{2}T\nu(\epsilon_{F})}{mR^{2}}\!\!\!\!\sum_{\mathbf{ia};\lambda_{2}\lambda_{3};m}\int_{\tau_{1}\tau_{2}}\!\!\!\!\! G_{\mathbf{i}\lambda_{2}}^{(0)}(\tau_{1},\tau_{2})\times\nonumber \\
\times\pmb\varphi_{\mathbf{i}1}(\tau_{1})\cdot\pmb\varphi_{\mathbf{i}1}(\tau_{2})G_{\mathbf{i}+\mathbf{a},\lambda_{3}}^{(0)}(\tau_{2},\tau_{1})\times\nonumber \\
\times\langle\Pi_{\mathbf{i},\mathbf{i}+\mathbf{a}}(\omega_{m})\rangle\text{sgn }(\omega_{m})\sin(\omega_{m}(\tau_{2}-\tau_{1})).\qquad\label{stunncorr6}\end{align}
 The right hand side of Eq.(\ref{stunncorr6}) vanishes because the
integrand is odd with respect to interchange of $\tau_{1}$ and $\tau_{2}.$
This proves that the correction to the effective action arising from
$\pmb\varphi_{\mathbf{i}1}$ fluctuations in $G_{\pmb\varphi_{\mathbf{i}1}}$
may be ignored.

\section{Optical conductivity of the AES model \label{sec:AEScond}}

We calculate the paramagnetic and diamagnetic terms in the AES conductivity
$\sigma_{\mathrm{AES}}$ of the granular array. As a corollary, we
also find that the resonance width $\Gamma$ is proportional to the
diamagnetic part of the AES conductivity. From Eq.(\ref{KAES}) through
Eq.(\ref{AEScurrent}) it follows that \begin{align}
\text{Re}(\sigma_{\mathrm{AES}}(\omega,T)) & =\frac{1}{\omega}\text{Im}(K_{\mathrm{AES}}^{dia}(\omega,T)+K_{\mathrm{AES}}^{para}(\omega,T)).\label{AEScond-appendix}\end{align}
 In the expression for $K_{\mathrm{AES}}^{dia}(\omega,T)$ we need
to calculate the cosine correlator, $\Pi(\tau)=\langle\Pi_{\mathbf{i},\mathbf{i}+\mathbf{e}_{\alpha}}(\tau)\rangle.$
We discuss the case of weak intergrain tunneling $g\ll1$ first.

\textbf{(a)} For weak intergrain tunneling, $\Pi(\tau)$ may be evaluated
perturbatively in increasing powers of $g:$ $\Pi(\tau)=\Pi^{(0)}(\tau)+\Pi^{(1)}(\tau)+\cdots,$
where the prefixes denote the power of $g.$ The leading term $\Pi^{(0)}(\tau)$
can be shown to be\cite{efetov02}\begin{align}
\Pi^{(0)}(\tau) & =\frac{1}{Z^{2}}\sum_{q_{1},q_{2}=-\infty}^{\infty}e^{-\beta E_{c}(q_{1}^{2}+q_{2}^{2})-(1-q_{1}-q_{2})2E_{c}\tau}.\label{Pi0def}\end{align}
 We similarly expand the diamagnetic response function $K_{\mathrm{AES}}^{dia}(i\Omega_{n})$
in powers of $g,$ $K_{\mathrm{AES}}^{dia}=K_{\mathrm{AES}}^{dia,(1)}+K_{\mathrm{AES}}^{dia,(2)}+\cdots,$
where the prefixes in brackets denote the power of $g.$ To obtain
the leading order in $g$ behavior we use Eq.(\ref{Pi0def}) in Eq.(\ref{KAESdia})
and take the Fourier transform:\begin{align}
K_{\mathrm{AES}}^{dia,(1)}(i\Omega_{n})=-g(e^{2}/a)\frac{T}{Z^{2}}\sum_{q_{1}q_{2}}\sum_{m}|\Omega_{m}|e^{-\beta E_{c}(q_{1}^{2}+q_{2}^{2})}\times\nonumber \\
\times(e^{-2E_{c}\beta(1-q_{1}-q_{2})}-1)\int d\Omega\,\delta(\Omega-2E_{c}(1-q_{1}-q_{2}))\times\nonumber \\
\times\left[\frac{1}{i\Omega_{n-m}-\Omega}-\frac{1}{i\Omega_{-m}-\Omega}\right].\label{AESdiafourier}\end{align}
 Next we perform the Matsubara sum over $m$ followed by an analytical
continuation $i\Omega_{n}\rightarrow\omega.$ The result is \begin{align}
\text{Im}(K_{\mathrm{AES}}^{dia,(1)}(\omega,T)) & =g(e^{2}/a)\frac{1}{Z^{2}}\sum_{q_{1}q_{2}}e^{-\beta E_{c}(q_{1}^{2}+q_{2}^{2})}\times\nonumber \\
\times\qquad\qquad & \!\!\!\!\!\!\!\!\!\!\!\!\!\!\!\!\!\!\!\!\!\!\!\int d\Omega\,(1-e^{-\beta\Omega})\frac{\Omega-\omega}{2}\left[\coth\frac{\Omega-\omega}{2}-\coth\frac{\Omega}{2}\right]\times\nonumber \\
\times & \delta(\Omega-2E_{c}(1-q_{1}-q_{2})).\label{AESdiafourier2}\end{align}
 The DC $(\omega=0)$ behavior in Eq.(\ref{AESdiafourier2}) is dominated
by single-charge excitations, $(q_{1},q_{2})=(1,0),\,(0,1),$ whereas
the a.c. behavior at $T=0$ is dominated by {}``even'' excitations
$(q_{1},q_{2})=(0,0),\,(1,1):$\begin{align}
\text{Im}(K_{\mathrm{AES}}^{dia,(1)}(0,T)) & \approx2\omega g(e^{2}/a)e^{-E_{c}/T},\nonumber \\
\text{Im}(K_{\mathrm{AES}}^{dia,(1)}(\omega,0)) & \approx\omega g(e^{2}/a)(1-2E_{c}/|\omega|)\times\nonumber \\
 & \times\Theta(|\omega|-2E_{c}).\label{AESdiafourier3}\end{align}
 In the next order in $g,$ the cosine correlator can be shown\cite{lohgranular05}
to be \begin{align}
\Pi^{(1)}(\tau) & =\frac{2\pi gT^{2}}{E_{c}^{2}\sin^{2}(\pi T\tau)},\,\, E_{c}\tau\gg1.\label{Pi1def}\end{align}
 It follows that in the second order in $g,$ the imaginary part of
the spectral function $K_{\mathrm{AES}}^{dia,(2)}$ is \begin{align}
\text{Im}(K_{\mathrm{AES}}^{dia,(2)}(\omega,T)) & =\omega\frac{4\pi g^{2}T^{2}e^{2}}{3aE_{c}^{2}}[T^{2}+(\omega/2\pi)^{2}],\,\omega\ll E_{c}.\label{AESdiafourier4}\end{align}
 The power law behavior of the second order (in $g$) diamagnetic
response does not mean that the conductivity will follow a power law.
This is because we also have a paramagnetic contribution, and one
can show\cite{lohgranular05} that the leading order paramagnetic
response is second order in $g$ and is \emph{equal and opposite}
to $K_{\mathrm{AES}}^{dia,(2)},$\begin{align}
K_{\mathrm{AES}}^{para}(\tau)\approx K_{\mathrm{AES}}^{para,(2)}(\tau) & =-K_{\mathrm{AES}}^{dia,(2)}(\tau).\label{cancellation}\end{align}
 Thus power law contributions cancel out in the conductivity and we
are left with \begin{align}
\sigma_{\mathrm{AES}}(\omega,T) & \approx\frac{\text{Im}(K_{\mathrm{AES}}^{dia,(1)}(\omega,T))}{\omega},\label{AESfinal}\end{align}
 where $\text{Im}(K_{\mathrm{AES}}^{dia,(1)}(\omega,T))$ is given
by Eq.(\ref{AESdiafourier2}).

\section{Behavior of the resonance width \label{sec:reswidth}}

Now we discuss the frequency and temperature dependence of the polarization
resonance width $\Gamma.$ Note that $\Gamma(\omega,T)$ is proportional
to $\sigma_{\mathrm{AES}}^{dia}(\omega,T).$ In the absence of a canceling
paramagnetic contribution, $\Gamma(\omega,T),$ unlike the conductivity,
does show an algebraic behavior at low frequencies, \begin{align}
\Gamma(\omega,T) & \approx4bzg^{2}\frac{4\pi e^{2}}{3RE_{c}^{2}}[T^{2}+(\omega/2\pi)^{2}],\omega\ll E_{c}.\label{gammaPowerlaw}\end{align}

Consider finally the case where the dimensionless intergrain tunneling
is large, $g\gg1.$ Except at very low temperatures (explained below),
both $\sigma_{\mathrm{AES}}$ and $\Gamma$ are more or less determined
by the diamagnetic contribution. Evaluating the cosine correlator,
\begin{align*}
\Pi(\tau) & \approx1-\frac{1}{\pi gz}\ln(gE_{c}\tau),\, gE_{c}\tau\gg1,\end{align*}
 and substituting in the expression for diamagnetic response, we have
\begin{align}
\sigma_{\mathrm{AES}}(\omega,T) & \approx g(e^{2}/a)\left[1-\frac{1}{\pi gz}\ln\left(\frac{gE_{c}}{\text{max}(\omega,T)}\right)\right].\label{AESlarge-g}\end{align}
 At exponentially small (in $g$) temperatures, such that the two
terms in the square brackets in Eq.(\ref{AESlarge-g}) become comparable,
perturbation theory in $1/g$ breaks down. Below such small temperatures,
the behavior of $\sigma_{\mathrm{AES}}$ (and $\Gamma$) is the same
as for the $g\ll1$ case, except that the charging energy $E_{c}$
in $g\ll1$ results should now be replaced with an effective charging
energy\cite{altland04,lohgranular05} $E^{*}(g)$ that is exponentially
small (in $g$) compared with $E_{c}.$


\end{document}